\documentclass[aps.pra,twocolumn,superscriptaddress,showpacs,fleqn,floatfix]{revtex4-1}
\usepackage{graphicx,amssymb}

\usepackage[fleqn]{amsmath}
\usepackage{amsfonts}
\usepackage{color}
\usepackage{braket}
\usepackage{indentfirst}
\usepackage{bm}
\usepackage{CJKutf8}

\renewcommand{\vec}[1]{\mbox{\boldmath $#1$}}

\usepackage{hyperref}

\begin{document}
\begin{CJK*}{UTF8}{gbsn}
\date{today}

\title{Resonant spectra of multipole-bound anions}

\author{X. Mao (毛兴泽)}
\affiliation{Department of Physics and Astronomy and FRIB Laboratory, Michigan State University, East Lansing, Michigan 48824, USA}

\author{K. Fossez}
\affiliation{NSCL Laboratory, Michigan State University, East Lansing, Michigan 48824, USA}

\author{W. Nazarewicz}
\affiliation{Department of Physics and Astronomy and FRIB Laboratory,
Michigan State University, East Lansing, Michigan 48824, USA}

\date{\today}

\begin{abstract}
In multipole-bound anions, the excess electron is attached by a short-range multipole potential of a neutral molecule. Such anions are prototypical marginally-bound open quantum systems. In particular, around the critical multipole moment required to attach the valence electron, multipole-bound anions exhibit critical behavior associated with a transition from bound states dominated by low-$\ell$ partial waves to the electron continuum. In this work, multipole-bound anions are described using a nonadiabatic electron-plus-rotor model. The electron-molecule pseudo-potential is represented by a short-range multipole field with a Gaussian form-factor. The resulting coupled-channel Schr\"odinger equation is solved by means of the Berggren expansion method, in which the electron's wave function is decomposed into bound states, narrow resonances, and the non-resonant scattering continuum. We show that the Gaussian model predicts the critical transition at the detachment threshold. Resonant states, including bound states, decaying resonances, subthreshold resonances, and antibound states are studied, and exceptional points where two resonant states coalesce are predicted. We discuss the transition of rotational band structures around the threshold and study the effects of channel coupling on the decay width of resonant poles.
\end{abstract}

\maketitle
\end{CJK*}

\section{Introduction}

The question of whether or not a neutral molecule can attach an excess electron to form a bound anion is not simple to answer \cite{desfrancois96_114,compton01_b60,jordan03_352,simons08_1079}. Fermi and Teller, in their pioneering work \cite{fermi47_110}, demonstrated the existence of the minimal dipole moment required to bind an electron in an external point dipolar field. This result stimulated many theoretical and experimental investigations on multipole-bound anions using effective potential methods \cite{desfrancois95_205,desfrancois96_114,abdoul98_108,abdoul02_1321,garrett70_118,garrett71_106,garrett82_104,ard09_122,desfrancois04_199,desfrancois98_1336,abdoul98_108,abdoul02_1321,fossez13_552,fossez15_1028,fossez16_1775} and ab-initio approaches \cite{jordan77_232,gutsev98_1097,adamowicz89_346,smith99_355,clary99_356,kalcher00_560,skurski02_575,peterson02_123,sommerfeld04_1573,sommerfeld14_1187,gutsev99_1324,gutsev98_1317,gutowski99_1186}. 

Because of the similarity of single-electron and rotational energy scales, there appears a strong, nonadiabatic coupling between the valence-electron and molecular rotational motions that impact the critical multipole moment required to form an anion \cite{herrick84_1248,clary88_345,clary89_303,brinkman93_208,ard09_122,garrett10_117}. Moreover, while the evidence for dipole-bound anions is solid, this is not the case for higher multipolarities \cite{simons08_1079,klahn98_752,fossez16_1775}.  In our previous study on the resonant spectrum of quadrupole-bound anions \cite{fossez16_1775} we predicted narrow resonances above the detachment threshold. The energies and widths of those resonances appear to be rather insensitive to details of the potential and are almost identical for prolate and oblate charge distributions.

While the binding of multipole-bound anions is fragile, low-energy resonances in such systems are expected to be less sensitive to details of the short-range molecular potential as the spatial extension of the valence electron is huge.
This situation resembles universal behavior, independent of the details of the interaction, exhibited by other weakly-bound/unbound quantum systems, such as nuclear and hadronic halos, cold atomic gases near a Feshbach resonance, and helium dimers and trimers, see, e.g., Refs.~\cite{Stecher2009,Hadizadeh2011,Hadizadeh2012,Lazauskas2013,Kievsky2014,konig17_1986,Deltuva2016,Deltuva2017,Shalchi2017,Miller2018}. In all of those cases, simple arguments based on scale separation and effective field theory capture the essential physics \cite{braaten06_823,bertulani02_869,bedaque03_1085,hammer10_1093,hammer00_1683,hammer17_1959,konig17_1986}. Consequently, to investigate generic properties of multipole-bound anions, we consider a schematic model, which contains the following crucial physics ingredients: (i) a short-range multipole potential and (ii) nonadiabatic coupling between electronic and molecular motion.

In the present study, we investigate the generic near-threshold behavior of multipole-bound anions at the transition between the subcritical and supercritical regimes. The main interest of this study is to show the role of low-$\ell$ partial waves in shaping the properties of low-lying states. We assume that the potential representing the molecular core is given by a Gaussian radial form-factor with a multipolar angular distribution. The particular choice of the radial form-factor is not important as it represents an a priori unknown short-range attraction. One can view this particular realization as a regularized zero-range interaction. The continuum couplings are included using the Berggren expansion method as in Refs.~\cite{fossez13_552,fossez15_1028,fossez16_1775}. 
To study the threshold behavior of the system we investigate the  pattern of resonant poles as a function of four parameters: the strength and range of the Gaussian form factor, the multipolarity of the potential, and the molecular moment of inertia. 

The paper is organized as follows. Section~\ref{s:mm} presents the model and method used. The results obtained in this study are discussed in Sec. \ref{s:rd}. Finally, Sec. \ref{s:conclusion} contains a summary and conclusions.

\section{Model and method} \label{s:mm}

\subsection{Hamiltonian}

\begin{figure}[t!]
	\includegraphics[width=0.7\linewidth]{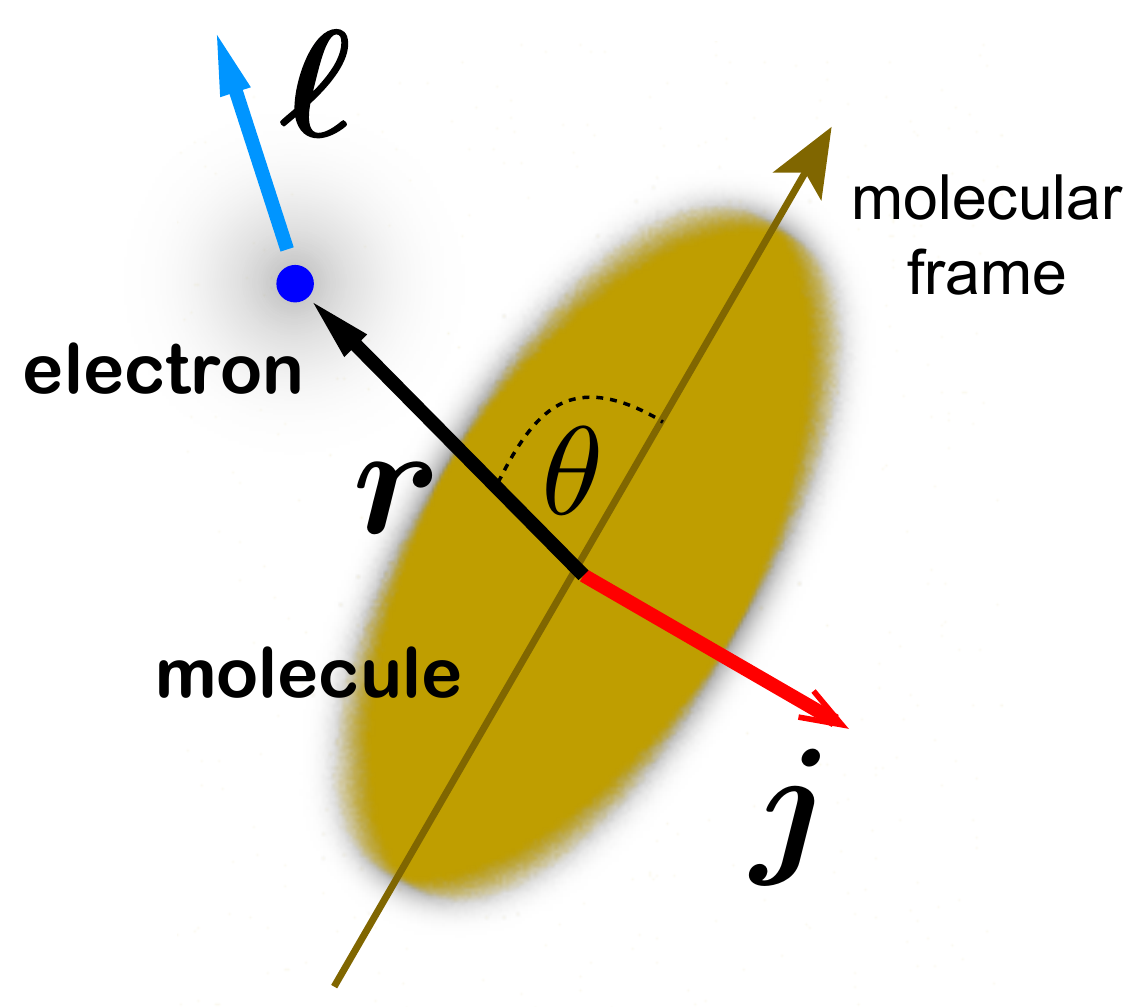}
	\caption{A schematic illustration of the electron-plus-molecule model used in this work.}
	\label{fig:model}
\end{figure}

In this work, we use the electron-plus-molecule Hamiltonian similar to that of Refs.~\cite{fossez13_552,fossez15_1028}. As shown in Fig.~\ref{fig:model} the valence electron is weakly coupled to the core. Without considering the spin-orbit interaction or the vibrational motion of the core, the Hamiltonian can be written as: 
\begin{equation}
	\hat{H}=\frac{\hat{j}^2}{2I} + \frac{\hat{p}_e^2}{2m_e} + V(\vec{r}).
\end{equation}
The first term is the rotational energy of the molecule with angular momentum $\hat{j}$ and moment of inertia $I$. The second term represents the kinetic energy of the electron of mass $m_e$ and linear momentum $\hat{p}_e$. The interaction between the rotor and the valence electron is modeled by the axially-deformed Gaussian potential of multipolarity $\lambda$: 
\begin{equation}
	V(\vec{r})=-V_0 \exp\left(-\frac{r^2}{2r_0^2}\right) P_{\lambda}(\cos\theta),
\end{equation}
where $\vec{r}$ is the electron's position vector in the molecular reference frame, $V_0$ is the potential strength, and $r_0$ is the potential range. The angular part of the potential is given by a Legendre polynomial of order $\lambda$, with $\theta$ being the angle between the direction of the valence electron $\hat{r}$ and the symmetry axis of the rotor. 

\subsection{Coupled-channel equations}

The total angular momentum of the system $\hat{J}$ is given by the sum of the angular momentum of the rotor $\hat{j}$ and the valence electron $\hat{\ell}$. 
Because the system is rotationally invariant in the laboratory reference frame, $\hat{J}$ commutes with the Hamiltonian $\hat{H}$ and the eigenvectors can be written as:
\begin{equation}
	\Psi^J=\sum_c u^J_c(r) \Theta_c^J
	\label{eq:wf}
\end{equation}
where $c$ labels all possible channels $(j,\ell)$ for a given $J$, $u^J_c(r)$ is the radial channel wave function, and $\Theta^J_c$ is the angular channel wave function. The eigenstates of Eq. \eqref{eq:wf} are also labeled by means for the parity quantum number $\pi$; hence, in the following we use the spectroscopic notation $J^\pi_n$, where $n=1$ marks the lowest $J^\pi$-state, $n=2$ - the next one, and so on. 
Due to the symmetries of $V(\vec{r})$, the ground-state rotational band of the molecule has states with $j=0,2,4,\dots$ and $\pi=+$ for $\lambda$-even  and 
$j^\pi=0^+,1^-,2^+,3^-,\dots$ for $\lambda$-odd \cite{Herzberg1}.

The coupled-channel equations are obtained by inserting the wave function \eqref{eq:wf} into the Schr\"odinger equation:
\begin{equation}
	\begin{aligned}
		&\left[\frac{d^2}{dr^2} - \frac{j_c(j_c+1 )}{I} - \frac{\ell _c(\ell_c +1)}{r^2} +E_J\right] u^J_c (r) \\
		&= \sum_{c^{\prime}} V^J_{cc^{\prime}} (r) u^J_{c^{\prime}} (r),
	\end{aligned}
	\label{eq:cc_eq}
\end{equation}
where $V^J_{cc^{\prime}} (r)$ is the channel-channel coupling potential. Throughout the paper, we will be using Rydberg units (energy expressed in Ry and distance in $a_0$). 

In the coupled-channel approach, the motion of the electron is weakly coupled to the rotation of the molecule. The adiabatic, or strongly-coupled, limit corresponds to an infinite moment of inertia where the rotational band of the molecule collapses to the band-head energy.

\subsection{Berggren expansion method}

The coupled-channel equations \eqref{eq:cc_eq} can be solved by means of the direct integration method (DIM), but a good initial guess is required to ensure convergence \cite{fossez16_1775}; this can be difficult for weakly bound states and broad resonances. Also, higher-multipolarity potentials require a larger number of channels, which makes this method computationally demanding.

\begin{figure}[tb]
	\includegraphics[width=0.9\linewidth]{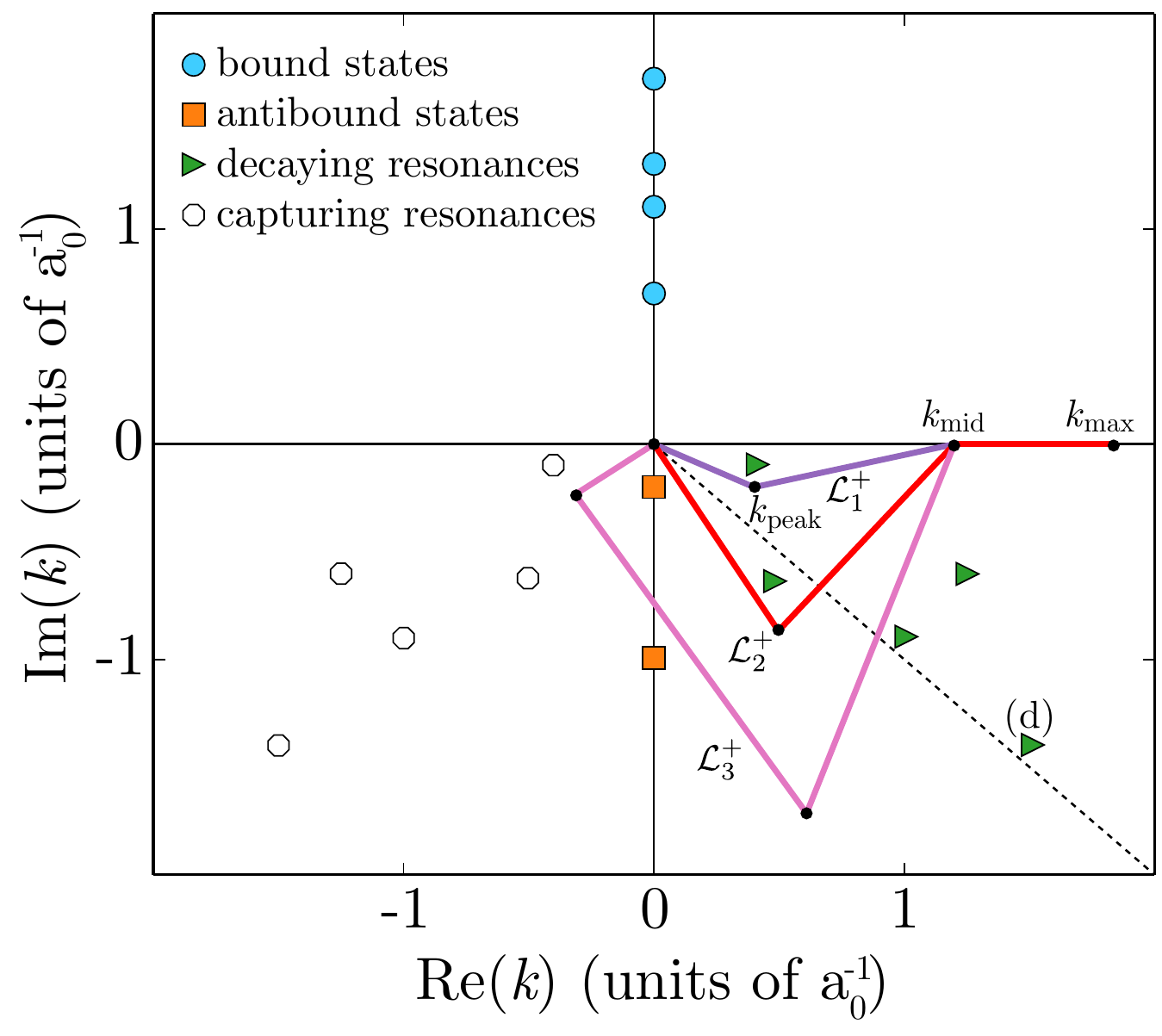}
	\caption{Berggren ensemble in the complex-$k$ plane. Bound, antibound, decaying, and capturing resonant states are marked. The distribution of poles is symmetric with respect to the imaginary $k$-axis because of time reversal symmetry. Three different scattering contours $\mathcal{L}_1^+$, $\mathcal{L}_2^+$, and $\mathcal{L}_3^+$ reveal $S$-matrix poles in different sectors in complex momentum/energy plane. The $-45^\circ$ line 
separating decaying resonances from subthreshold resonances is marked.}
\label{figcomplexk}
\end{figure}

An alternative to the DIM is the Berggren expansion method (BEM), previously applied in the context of multipole-bound anions \cite{fossez15_1028,fossez16_1775} and nuclear halos \cite{hagen06_464,papadimitriou11_277,fossez16_1335}. The Berggren basis \cite{berggren68_32,berggren93_481} used in this work is defined in the complex momentum plane; it contains explicitly resonant states (poles of the one-body $S$-matrix) and scattering states defined along a contour $\mathcal{L}^+$ in the fourth quadrant of the momentum plane. The completeness relation for the Berggren ensemble can be written as:
\begin{equation}
	\sum_{b,a,d} \ket{\tilde{u}_n} \bra{u_n} + \int_{\mathcal{L}^+} \ket{\tilde{u} (k)} \bra{u(k)} dk =1,
	\label{eqBerggren}
\end{equation}
where the sum over discrete resonant states includes bound states $b$, antibound (or virtual) states $a$, and decaying poles $d$ lying between the positive real axis and the contour $\mathcal{L}^+$. The tilde symbol indicates time reversal. In the unlikely situation that bound states of energies higher than antibound states are present, they must be excluded from the sum in~Eq.~(\ref{eqBerggren}). The decaying poles in the fourth quadrant, which lie close to the real $k$-axis and have a real energy Re$(E)>0$ and a width $\Gamma$=$-$2Im$(E)>0$ can be interpreted as narrow resonances. The poles with Re$(E)<0$ and $\Gamma>0$, located below the $-45^\circ$ line in Fig.~\ref{figcomplexk} and close to the origin, can be associated with subthreshold resonances \cite{Kok1980,mukhamedzhanov10_210,Mukhamedzhanov2017,Sofianos1997}. 

In practical applications, one often considers a contour $\mathcal{L}_1^+$ of Fig.~\ref{figcomplexk} that starts at the origin, extends into the fourth quadrant up to $k_{\text{peak}}$, comes back to the real axis at $k_{\text{mid}}$, and continues along the real axis up to the cutoff momentum $k_{\text{max}}$. To be able to explore $S$-matrix poles in other regions of the complex momentum plane, two other contours are used in this work. With the contour $\mathcal{L}_2^+$ we explore the region of subthreshold resonances. The contour $\mathcal{L}_3^+$ can be employed to reveal antibound states lying on the negative imaginary momentum axis and the capturing resonances lying in the third complex-$k$ quadrant.

For each channel, the basis is generated using the diagonal part of the potential in the channel basis $V_{cc}$ \cite{fossez15_1028}. Bound states and decaying resonances entering the Berggren basis for a given partial wave are obtained by a direct integration of the Schr\"odinger equation for the diagonal term of the potential, while the selected scattering states along the contour $\mathcal{L}^+$ are discretized in the momentum space using a Gauss-Legendre quadrature as in Refs.~\cite{fossez15_1028,fossez16_1775}. The non-resonant continuum is limited by the momentum cutoff $k_{\text{max}}$ that has to be sufficiently large to ensure the completeness of the Berggren basis. While the bound states are normalized in the standard way, decaying resonances are normalized using the exterior complex scaling method \cite{gyarmati71_38,simon79_436}. The scattering states are normalized to Dirac-delta function. This representation provides a natural way to include continuum couplings for each desired partial wave. 

The spectrum of the system is obtained by diagonalizing the complex-symmetric Hamiltonian matrix. It consists of resonant eigenstates representing bound states and narrow resonances, and non-resonant scattering solutions. Differentiating resonant states from the non-resonant scattering background requires special treatment. Since resonant states do not depend on a detailed choice of the contour $\mathcal{L}^+$, by moving the contour slightly a new spectrum can be obtained, where non-resonant states move according to the contour change and resonant states stay invariant \cite{fossez16_1775}. In this way, resonant states can be located. As a further test, these resonant states are used in the DIM as an initial guess, and it is checked that the BEM results are reproduced.

\section{Results} \label{s:rd}

\subsection{Threshold trajectories for multipolar Gaussian potentials in the adiabatic limit \label{sub:threshold_line}}

\begin{figure}[htb]
	\includegraphics[width=0.9\linewidth]{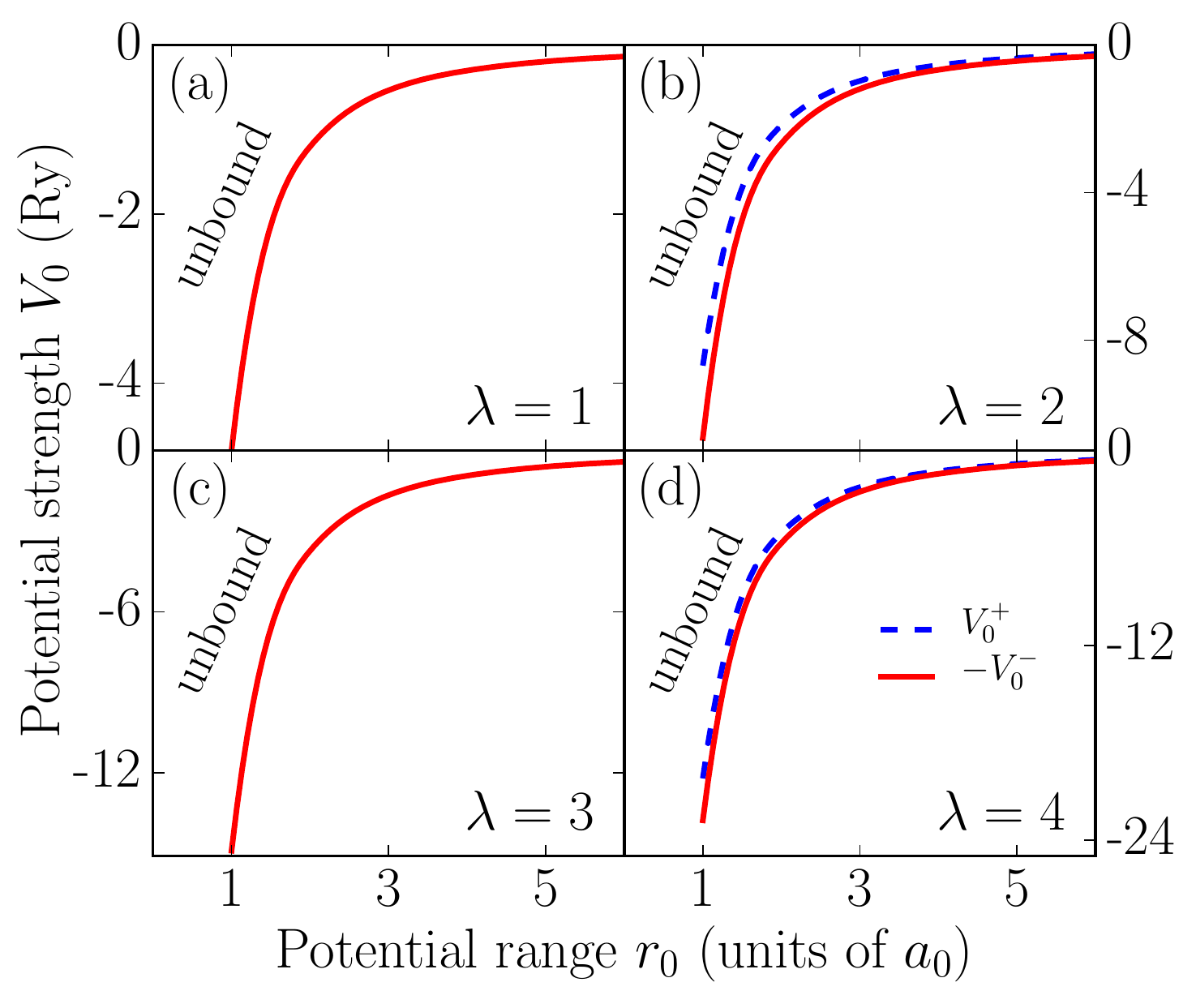}
	\caption{Threshold trajectories $({V}_{0}, r_0)_{c}^{\pm}$ for multipolar Gaussian potentials with $\lambda=1-4$ in the adiabatic limit.}
	\label{fig:threshold_line}
\end{figure}

Multipole-bound anions can be characterized by their critical multipole moments $Q_{\lambda,c}^\pm$, which mark the limit between the subcritical and supercritical regimes. 
We note that for odd-multipole potentials $Q_{\lambda,c}^-=-Q_{\lambda,c}^+$, but there is no such relation for even-multipole potentials. For instance, there are two critical values of the quadrupole moment for a quadrupole-bound anion ($\lambda=2$): $Q_{2,c}^+$ (prolate) and $Q_{2,c}^-$ (oblate), and $Q_{2,c}^- \ne -Q_{2,c}^+$.

As the usual ${ -1/{r}^{\lambda+1} }$ radial dependence of multipolar potentials is replaced in our work by the Gaussian form factor, the detachment threshold is obtained at the critical trajectories of $({V}_{0}, r_0)_{c}^{\pm}$. Figure~\ref{fig:threshold_line} shows such trajectories obtained in the adiabatic limit for the $J^{\pi}=0^+_1$ ground states of anions with multipolarities $\lambda=1-4$.

The complex-momentum contours used in the Berggren basis are defined by the points $k=(0,0), k_{\text{peak}}=(0.5,-0.1)$, $k_{\text{mid}}=1.0$, and $k_{\text{max}}=14.0$ (in units of ${ {a}_{0}^{-1} }$), with each segment being discretized by 40 Gauss-Legendre points. To ensure convergence, we took $\ell_{\text{max}}=4$ for $\lambda=1,2,3$ and $\ell_{\text{max}}=8$ for $\lambda=4,5$. 

As one would expect, the absolute value of the critical potential strength ${ | {V}_{0,c} | }$ required to bind an excess electron is decreasing with the range $r_0$ and for a fixed range ${ | {V}_{0,c} | }$ increases with multipolarity. Also, as noted in previous studies \cite{neirotti97_1466,ferron04_130,pupyshev04_161,fossez16_1775}, for even multipolarities, the value of ${ | {V}_{0,c} | }$ for negative-$V_0$ potentials (``prolate") is larger than that for positive-$V_0$ potentials (``oblate"). 

It is interesting to note that at the threshold, the wave functions are dominated by the $\ell=0$ component. Dividing the intrinsic wave function into the inner region $(r < R)$ and outer region $(r > R)$ contributions, where $R$ is the distance at which the core potential becomes practically unimportant, one can show \cite{misu97_1181,riisager92_615,yoshida05_1895} that the probability of finding the electron in the outer region approaches one at the detachment threshold, if the $\ell=0$ component is present in the intrinsic wave function. This has been practically demonstrated in our previous work on quadrupole-bound anions \cite{fossez16_1775} in the context of the scaling properties of root-mean-square radii.

\subsection{Resonances of the near-critical quadrupolar Gaussian potential \label{sub:resonances}}

In order to study the role of low-$\ell$ partial waves in multipole-bound anions at the interface between the subcritical and supercritical regimes, one has to recognize the impact of $\ell = 0$ partial waves on resonant states near threshold \cite{yoshida05_1895}. In our coupled-channel formalism, resonant states appear through the mixing of different channels. To study general features of near-threshold resonances, we consider three states of the quadrupolar potential in the adiabatic approximation. Namely, we investigate: (i) the ${J}^{\pi} = {0}_{1}^{+}$ ground state dominated by the $\ell = 0$ partial wave; (ii) an excited ${ {J}^{\pi} = {0}_{d}^{+} }$ state dominated by the $\ell=2$ channel; and (iii) the lowest ${ {J}^{\pi} = {1}_{1}^{-} }$ state, which is primarily $\ell = 1$. The quadrupolar case discussed here is characteristic of other multipolar potentials.

\subsubsection{Resonant states dominated by the $\ell=0$ channel}\label{swave} 

The ground state (g.s.) of the quadrupolar potential is computed with the BEM, using the extended contour $\mathcal{L}_3$ of Fig.~\ref{figcomplexk} defined by the points: $k=(0,0)$, $(-0.1,-0.4)$, $(0.1,-0.4)$, $(2,0)$, and $(14,0)$ (all in $a_0^{-1} $), each segment being discretized with 40 Gauss-Legendre points. By considering the contour that extends into the third quadrant of the complex momentum plane, antibound states can be revealed, see Fig.~\ref{figcomplexk} and Refs.~\cite{betan04_37,michel06_16,michel09_2}. 

\begin{figure}[htb]
	\includegraphics[width=0.9\linewidth]{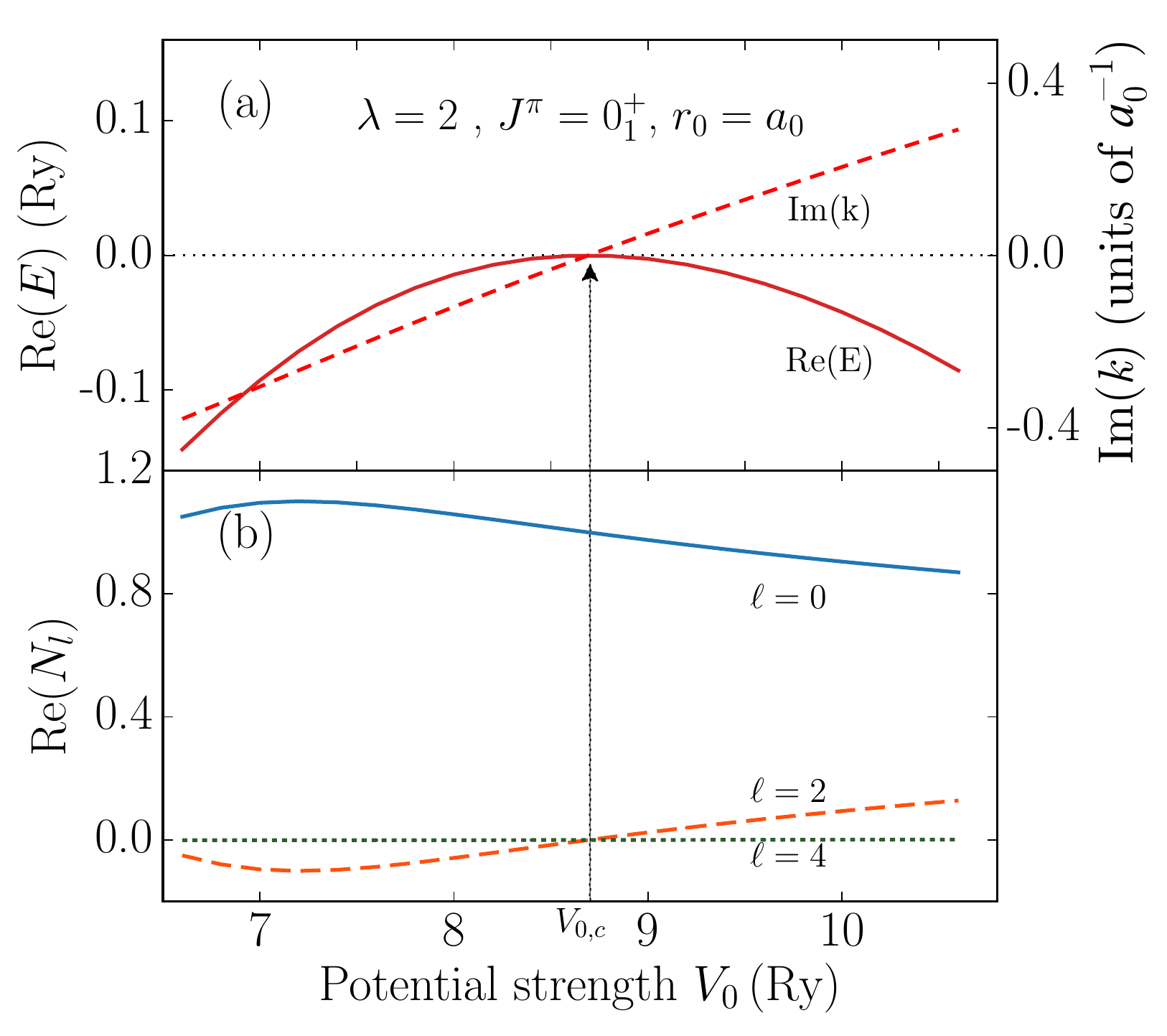}
	\caption{The lowest $0^+$ resonant state of the quadrupolar Gaussian potential with 
		$ r_0 = a_0 $ as a function of $V_0$. Top: real energy and imaginary momentum. Bottom: the channel decomposition of the real part of the norm. The critical strength ${V}_{0,c}$ is marked by arrow.
	}
	\label{fig:antibound}
\end{figure}

Figure~\ref{fig:antibound}(a) shows the energy and momentum of the ${0}_{1}^{+}$ state for different values of the potential strength $V_0$. For large values of $V_0$, the g.s. is bound (Re($E$)$<$0) and has a positive imaginary momentum. As the potential strength decreases, the energy of the ground state moves up and approaches the $E=0$ threshold at ${V}_{0,c} = 8.7$\,Ry. For $V_0<{V}_{0,c}$ the lowest ${0}^{+}$ state becomes antibound (Re($E)<0$, Im$(k)<0$). As illustrated in Fig.~\ref{fig:antibound}(b), the contributions ${ {N}_{\ell} }$ to the complex norm of the wave function from different $\ell$-channels ($\ell = 0, 2$, 4) vary smoothly when crossing the threshold. The norm is largely dominated by the $\ell = 0$ component. At the critical strength, the $\ell>0$ contributions to the norm vanish, cf. discussion in Sec.~\ref{sub:threshold_line}. The presence of near-threshold antibound states impacts the structure of the low-energy continuum and can manifest their existence through peaks in the scattering cross section at low-energy \cite{rohr75_283,rohr76_284,rohr78_1426,heiss11_776,mukhamedzhanov10_210}.

\subsubsection{Resonant states dominated by  a $\ell \neq 0$ channel}

\begin{figure}[htb]
	\includegraphics[width=0.9\linewidth]{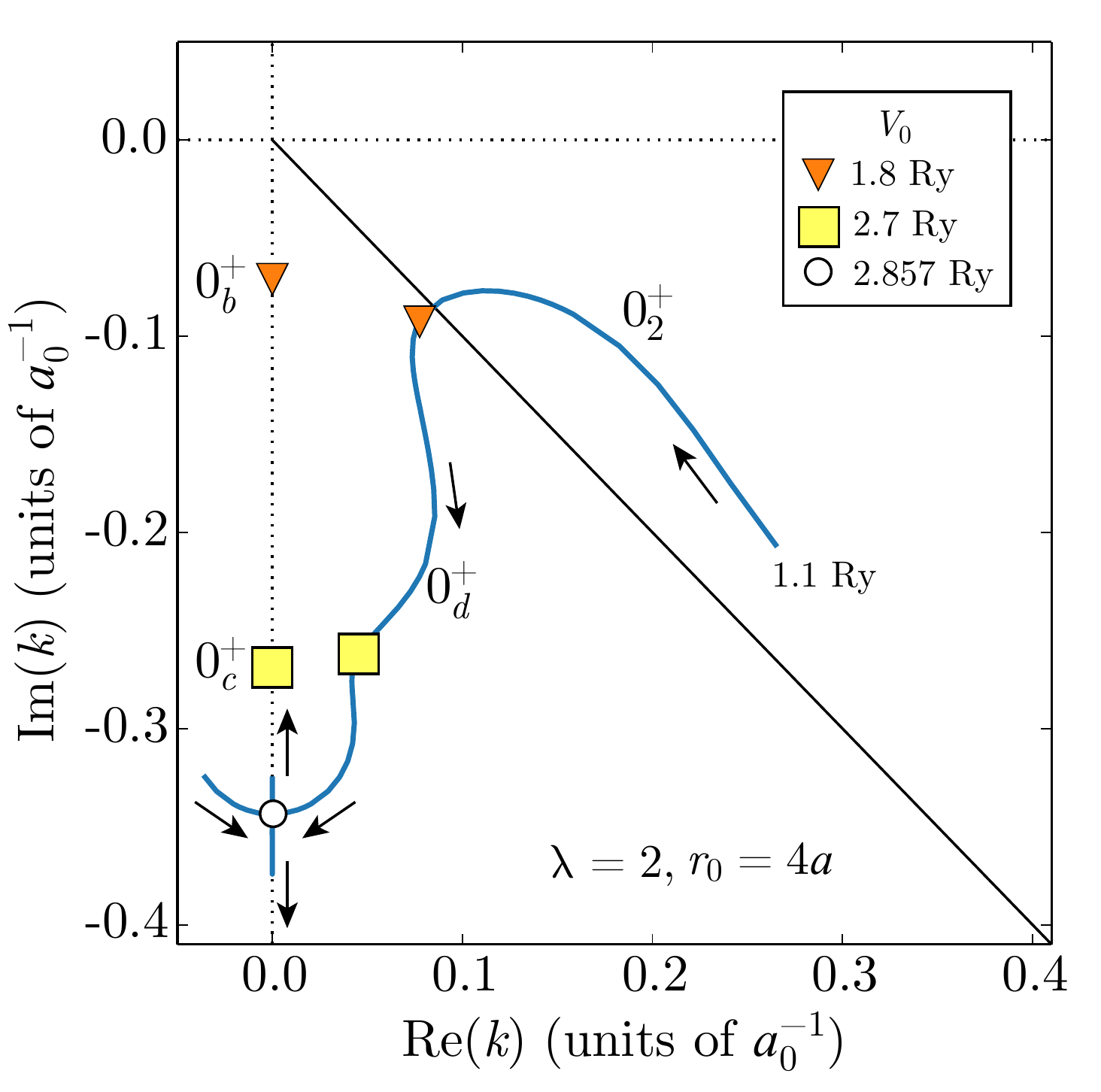}
	\caption{Trajectory of the $0^+$ resonant state in the complex-$k$ plane of the quadrupolar potential with 
	 $r_0 = 4\,a_0 $ as the 
	 potential strength $V_0$ increases in the direction indicated by an arrow. At the lowest value $V_0$ =1.1\,Ry, the $0^+$ ground state is bound and the state of interest is an excited ${0}_2^+$ state associated with a decaying resonance. At $V_0=1.8$\,Ry the pole crosses the $-45^\circ$ line and becomes a subthreshold resonance $0_d^+$. At $V_0=2.857$\,Ry the decaying pole reaches the imaginary-$k$ axis and coalesces with the capturing pole with Im$(k)<0$ forming an exceptional point. The antibound states at $V_0=1.8\,$Ry and 
	$V_0=2.7$\,Ry are marked.}
	\label{fig:2nd_resonant}
\end{figure}

\begin{figure}[htb]
	\includegraphics[width=0.9\linewidth]{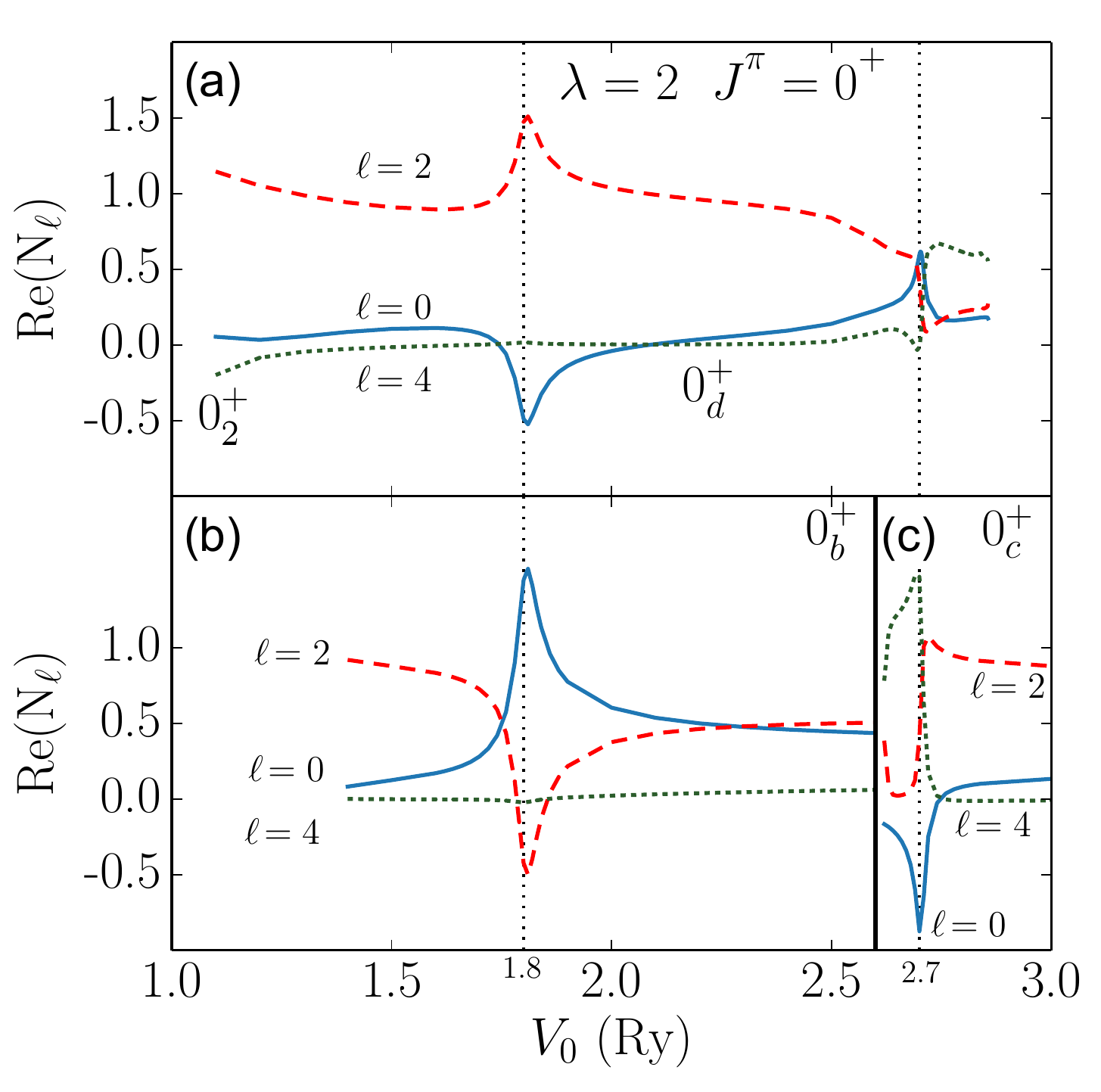}
	\caption{Real norms of the channel wave functions for the decaying pole $0_d^+$ shown in Fig.~\ref{fig:2nd_resonant} and the antibound states $0_b^+$ and $0_c^+$ of Fig.~\ref{fig:antibound_transition}.}
	\label{fig:0_2_norm}
\end{figure}

We now consider the evolution of an excited state of the quadrupolar potential with $ r_0 = 4 \, a_0 $. At $V_0$ =1.1\,Ry the lowest $0^+$ state is bound and the second ${J}^{\pi} = {0}_2^+$ state is a decaying resonance, see Fig.~\ref{fig:2nd_resonant}. Figure \ref{fig:0_2_norm}(a) shows the channel decomposition for this state. It is seen that its configuration has the predominant $\ell=2$ component.

As the potential gets deeper, the pole crosses the $-45^\circ$ line at $V_0 \approx 1.8$\,Ry and becomes a subthreshold resonance labeled as $0_d^+$. At $V_0=2.7$\,Ry a rapid transition to a configuration dominated by the $\ell=4$ partial wave takes place, which is indicative of a level crossing in the complex-$k$ plane. At $V_0=2.857$\,Ry the decaying pole arrives at the imaginary-$k$ axis and coalesces with the symmetric capturing pole forming an exceptional point \cite{heiss12_1392,Muller2008_1973,Okolowicz2009_1974}. At still larger values of $V_0$, the exceptional point splits up into two antibound states moving up and down along the imaginary $k$-axis as shown in Fig.~\ref{fig:2nd_resonant}. A similar situation was discussed in Refs.~\cite{domcke81_289,Garmon2015_1975} in the context of electron-molecule scattering and optical lattice arrays, respectively.

\begin{figure}[htb]
	\includegraphics[width=0.9\linewidth]{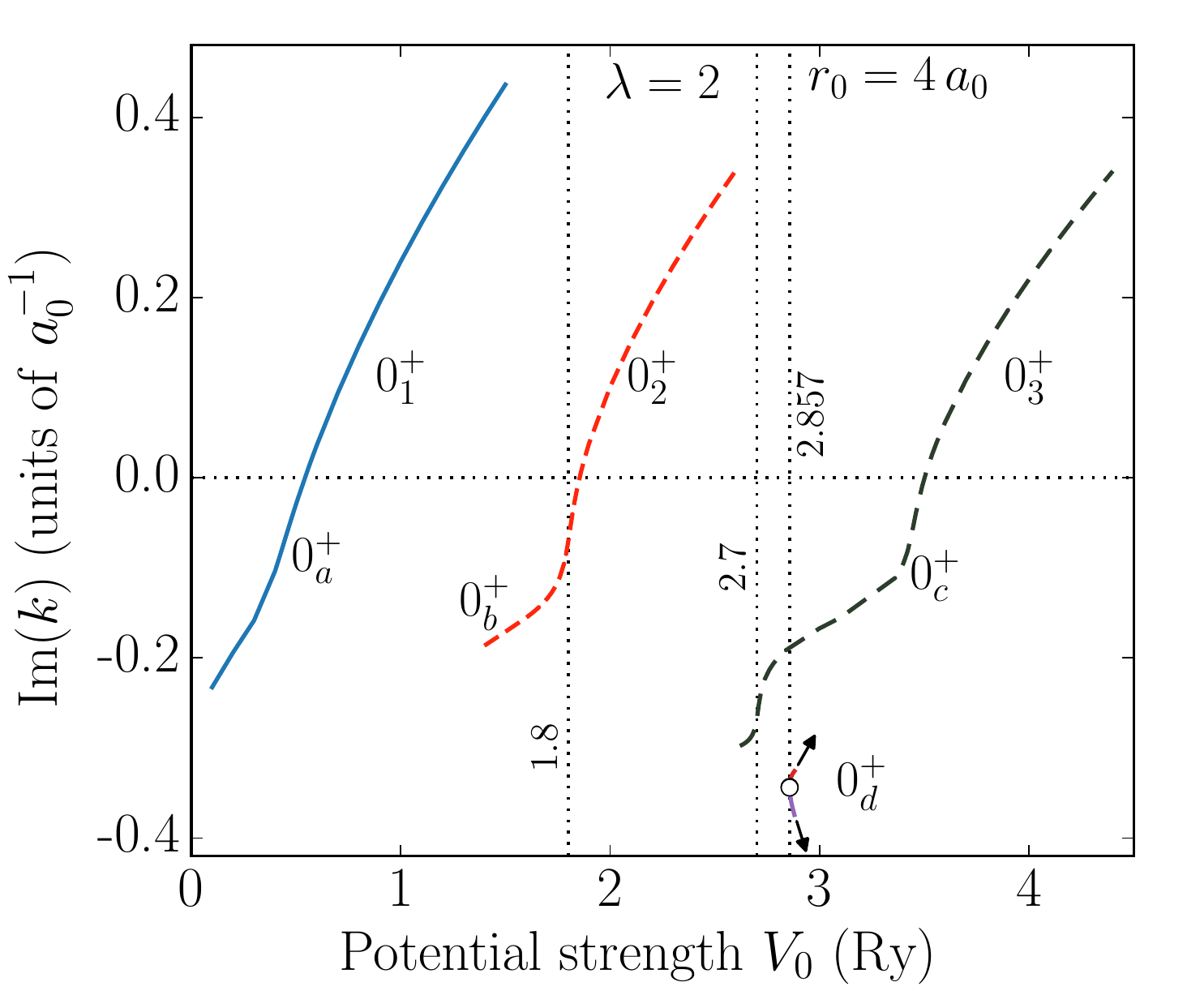}
	\caption{Trajectories of antibound and bound $0^+$ states along the imaginary $k$-axis as a function of $V_0$ for the quadrupolar potential with $r_0 = 4\,a_0$. With increasing potential strength, the antibound states $0_a^+$, $0_b^+$, and $0_c^+$ become bound states of the system $0_1^+$, $0_2^+$, and $0_3^+$, respectively. The open circle marks the exceptional point of Fig.~\ref{fig:2nd_resonant}, which is the source of two antibound states. The particular values of $V_0$ discussed around Fig.~\ref{fig:2nd_resonant} are marked.}
	\label{fig:antibound_transition}
\end{figure}

In the range of $V_0$ corresponding to the trajectory $0^+_2 \rightarrow 0^+_d$ shown in Fig.~\ref{fig:2nd_resonant}, there appear antibound states in the threshold region. Their trajectories along the imaginary $k$-axis are shown in Fig.~\ref{fig:antibound_transition} and their channel decompositions are given in Fig.~\ref{fig:0_2_norm}(b) and (c). As $V_0$ increases, the antibound states $0_a^+$, $0_b^+$, and $0_c^+$ emerge as bound physical states of the system labeled as $0_1^+$, $0_2^+$, and $0_3^+$, respectively. The lowest antibound state $0^+_a$ has a dominant $\ell=0$ configuration, similar to that of Fig.~\ref{fig:antibound}. At low values of $V_0$, the wave function of the antibound state $0^+_b$ is predominantly $\ell=2$. As seen in Fig.~\ref{fig:2nd_resonant}, this state appears close to the decaying pole $0^+_d$ at $V_0 \approx 1.8$\,Ry and the crossing between these two poles in the complex-$k$ plane is seen in their wave function decompositions. Following the crossing, the state $0^+_b$ acquires a large  $\ell=0$ component. The antibound state $0^+_c$ begins as an $\ell=4$ configuration. At $V_0\approx 2.7$\,Ry, this state interacts with $0^+_d$ and its configuration changes to $\ell=2$. One can thus see that the presence of antibound states results in the particular shape of the $0^+_d$-pole trajectory in the complex-$k$ plane.

\begin{figure}[htb]
	\includegraphics[width=0.9\linewidth]{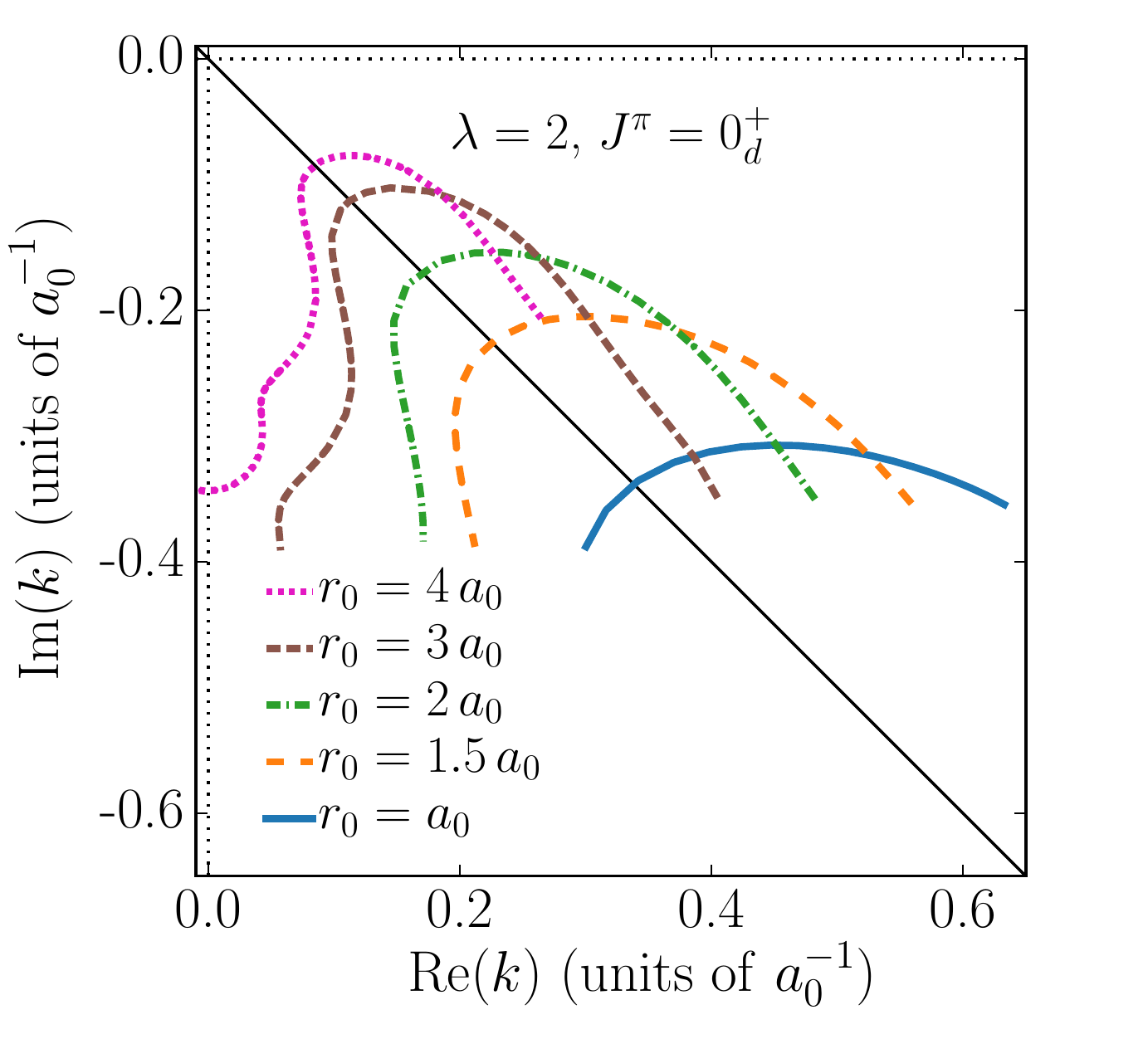}
	\caption{Trajectory of the $0_d^+$ resonant state in the complex-$k$ plane for different values $r_0$ of quadrupolar potential as indicated by numbers (in units of $a_0$). The ranges of $V_0$ (in Ry) are: (25.6-29.0) for $r_0=a_0$;
		(9.7-14.5) for $r_0=1.5\,a_0$;
		(4.8-10) for $r_0=2\,a_0$;
		(1.7-4.79) for $r_0=3\,a_0$; and
	(1.1-2.85) for $r_0=4\,a_0$.}
	\label{fig:0+_dif_c}
\end{figure}

The dependence of the $0^+_d$-pole trajectory on the potential range is illustrated in Fig.~\ref{fig:0+_dif_c}.
For potentials with longer ranges, pole trajectories appear closer to the origin. In all the cases shown, a transition from decaying to subthreshold resonances takes place. These poles have large widths, and are expected to impact the structure of the low-energy scattering continuum.

\subsubsection{Resonant states without a $\ell = 0$ component}

Here we discuss the lowest ${ {J}^{\pi} = {1}_{1}^{-} }$ state, which is primarily $\ell = 1$ with a small admixture of the $\ell = 3$ channel. This case closely follows the discussion of Ref.~\cite{domcke81_289} for $p$-wave scattering from short-range potentials.
\begin{figure}[htb]
	\includegraphics[width=0.8\linewidth]{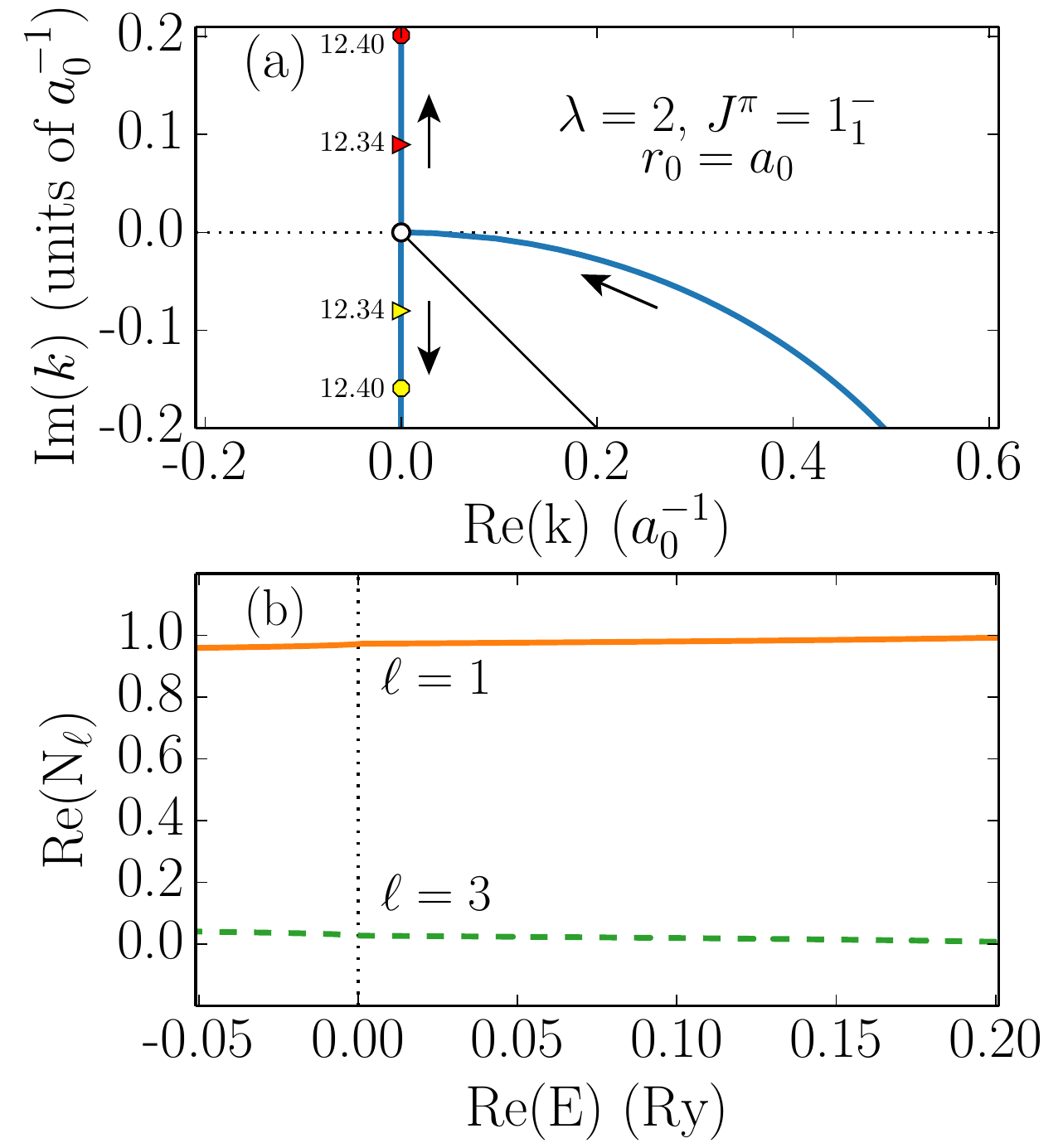}
	\caption{Top: trajectory of the lowest $1_1^-$ resonant state of the quadrupolar potential with $r_0=a_0$ as a function of $V_0$ in the range of (9-12.7)\,Ry.
		The potential strength $V_0$ increases along the direction  indicated by an arrow. The positions of the bound and antibound states at $V_0=12.34$\,Ry and 12.4\,Ry are marked.
		Bottom: real norms of channel functions for this state.
	}
	\label{fig:1-_resonant}
\end{figure}
The corresponding trajectory of this state in the complex momentum plane is shown in Fig.~\ref{fig:1-_resonant}(a). At larger values of $V_0$, the $1_1^-$ state is bound. As $V_0$ decreases, this state crosses the detachment threshold and becomes a narrow decaying resonance. The trajectory of the capturing resonance, symmetric with respect to the Im$(k)$ axis, is not shown. As discussed in Ref.~\cite{domcke81_289}, the exceptional point appears at the origin at $V_{0,c}$. Close to the threshold, the bound state and the antibound state are located symmetrically to the origin. For the $p$-wave dominated state, the transition from the subcritical to the supercritical regime is smooth, i.e., the wave function amplitudes hardly change with $V_0$, see Fig.~\ref{fig:1-_resonant}(b). This is because the contributions from antibound and bound state poles cancel each other out. In this case, the structure of the low-energy continuum is not expected to be affected by the presence of threshold poles.

The situation presented in Fig.~\ref{fig:1-_resonant} is rather generic for  $p$-wave dominated resonant poles. Increasing the potential range moves the pole trajectory closer to the real-$k$ axis. Consequently, states containing no $s$-wave component are  likely to appear as isolated narrow resonances. 
For odd-multipolarity potentials, a ($j=J, \ell=0$) component of a  $J^\pi$ state  becomes large as the detachment threshold is approached, see Sec.~\ref{swave}. On the other hand, for even-multipolarity potentials,  odd-$J$ states cannot have an $s$-wave component, as the core's angular momentum $j$ must be even, and narrow near-threshold resonances can appear.

 \subsection{Rotational motion}\label{sub:rotational}

To describe multipole-bound anions, one has to  take into account the nonadiabatic coupling between the rotational motion of the molecule and the single-particle motion of the electron. Whether a multipole-bound anion can exhibit rotational bands depends on the molecule's multipolarity. For instance, it was shown in Ref.~\cite{fossez15_1028} that rotational bands of dipolar anions do not extend above the detachment threshold while a similar study for quadrupole-bound anions \cite{fossez16_1775} demonstrated that the rotational motion of the anion is hardly affected by the continuum effects. The reason for this difference might be due to the existence of two coupling regimes in the dipolar case: a strong coupling regime below the threshold (valence electron follows the rotational motion of the core) and a weak coupling regime in the continuum region (valence electron is almost entirely decoupled from the molecular rotation).

\begin{figure}[htb]
	\includegraphics[width=0.8\linewidth]{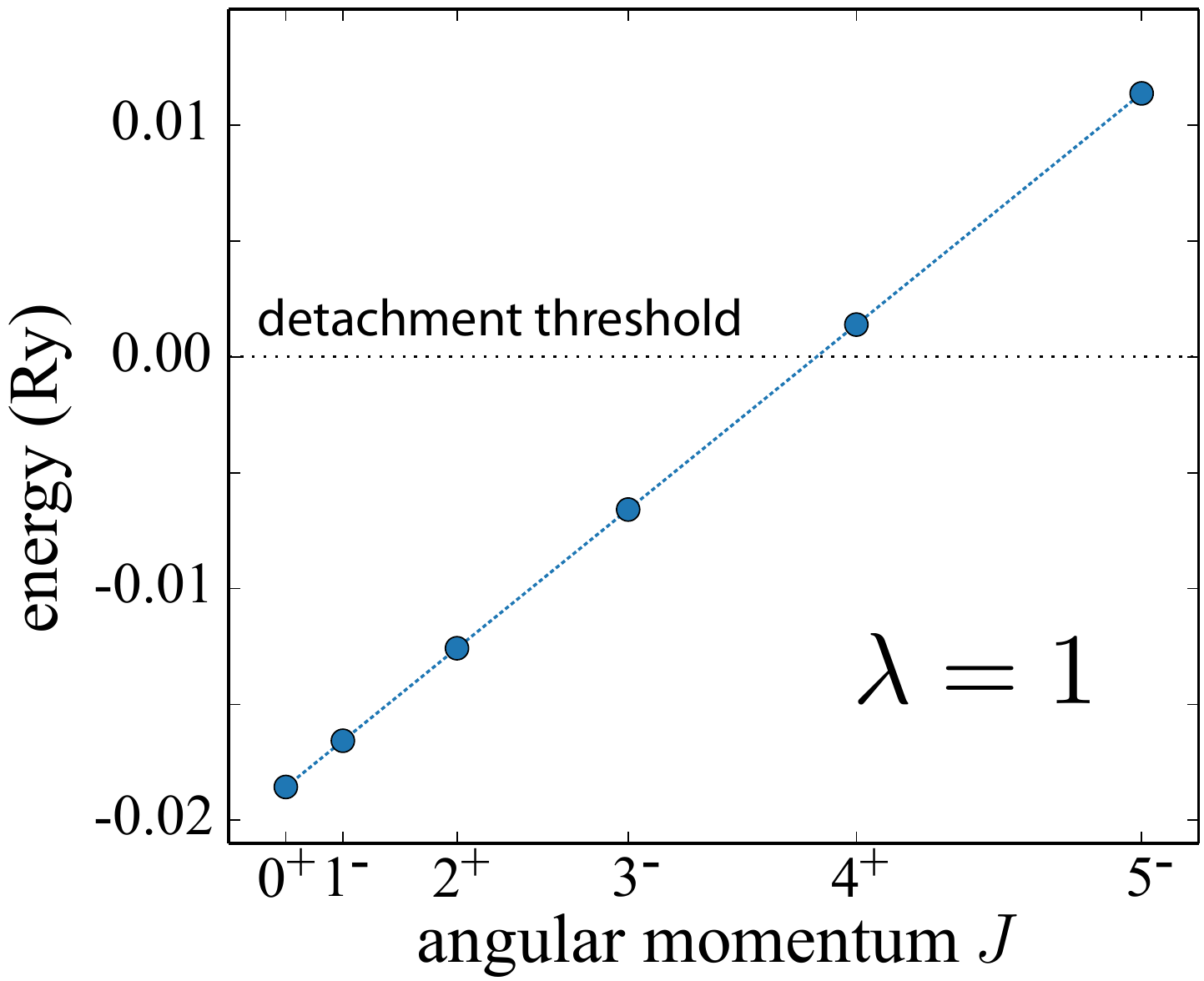}
	\caption{The rotational band built upon the 
		${ {J}^{\pi} = {0_1^+} }$ state of a dipole-bound anion.
		The parameters $V_0= 5.33$\,Ry, $r_0=a_0$, and $I=10^{3} \, m_e a_0^2$
		have been chosen to place the band-head energy slightly below the zero-energy  threshold, where rotational motion of the rotor can excite the system into the continuum.
		The energy is plotted as a function of $J(J+1)$. 
	}
	\label{fig:lambda_1_rota_band}
\end{figure}
Figure~\ref{fig:lambda_1_rota_band} illustrates the case of a rotational band built upon the subthreshold ${ {J}^{\pi} = 0_1^+ }$ state of the Gaussian dipolar potential. It is seen that the rotational band is not affected when the zero-energy threshold is crossed below $J=4$. This result indicates that the presence of the two coupling regimes predicted to exist in realistic calculations for dipole-bound anions~\cite{fossez15_1028} must be due to difficulties in imposing proper boundary conditions at infinity for the dipolar potential ($\sim r^{-2}$)  when the rotational motion of the molecule is considered nonadiabatically \cite{fossez13_552}. Since in the present work the radial part of the dipolar pseudopotential is replaced by a Gaussian,  the outgoing boundary condition can be readily imposed.

\begin{figure}[htb]
	\includegraphics[width=0.8\linewidth]{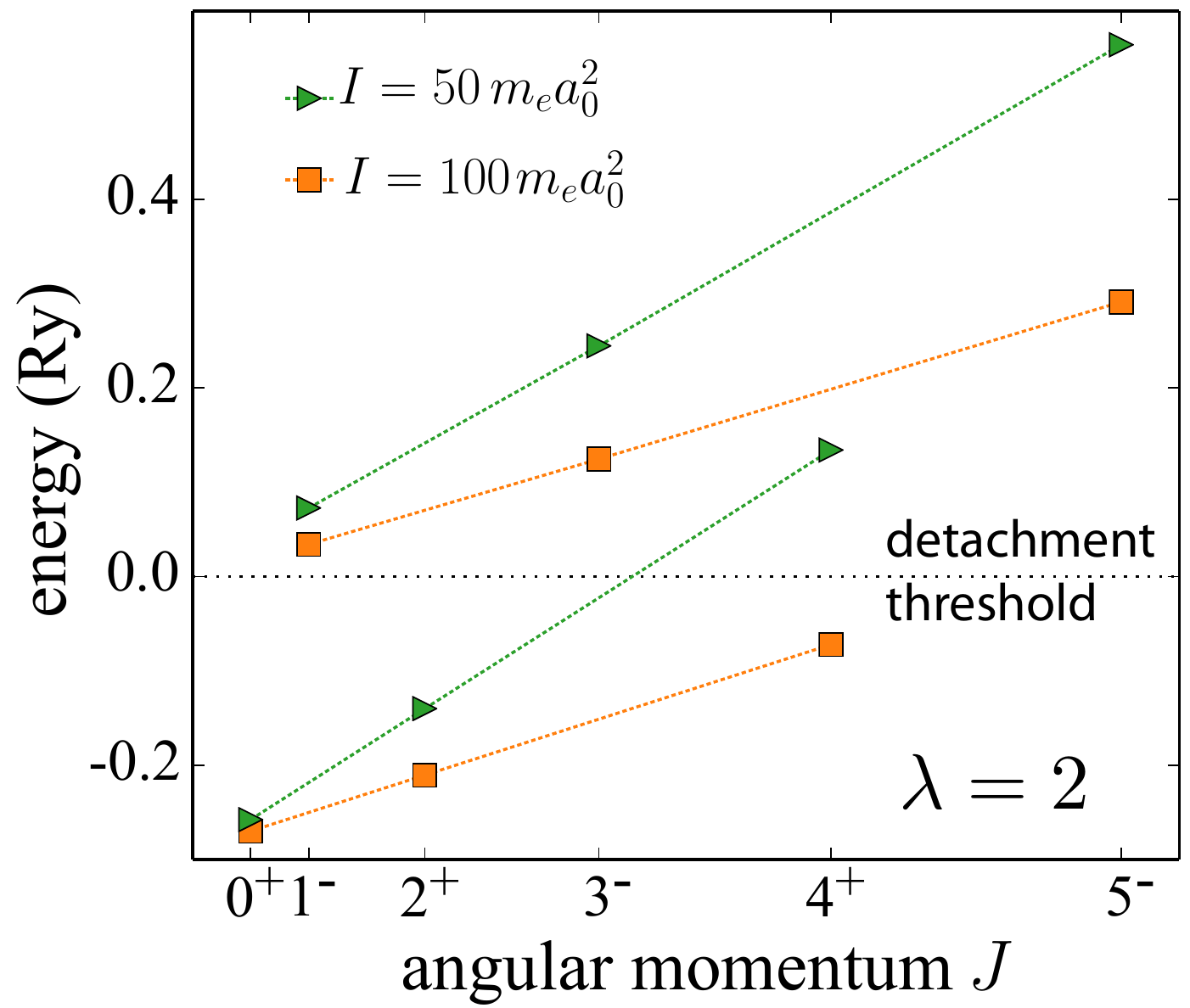}
	\caption{Similar to Fig.~\ref{fig:lambda_1_rota_band} but for rotational bands built upon the ${ {J}^{\pi} = {0_1}^{+} }$ and ${1_1}^{-}$ bandheads of a quadrupolar Gaussian potential with $V_0=12.38$\,Ry, $r_0=a_0$, and for $I=50\, m_e a_0^2$ and $I=100\, m_e a_0^2$.}
	\label{fig:lambda_2_rota_band}
\end{figure}
A similar result is obtained for the quadrupolar case shown in Fig.~\ref{fig:lambda_2_rota_band} for two rotational bands built upon the ${ {J}^{\pi} = 0_1^+ }$ and $1_1^-$ bandheads. The existence of rotational bands extending above the detachment threshold is consistent with the findings of Ref.~\cite{fossez16_1775} employing the realistic quadrupolar pseudopotential. The results for higher-multipolarity potentials follow the pattern obtained for the dipolar and quadrupolar cases; hence, they are not shown here.

\begin{figure}[htb]
	\includegraphics[width=1.0\linewidth]{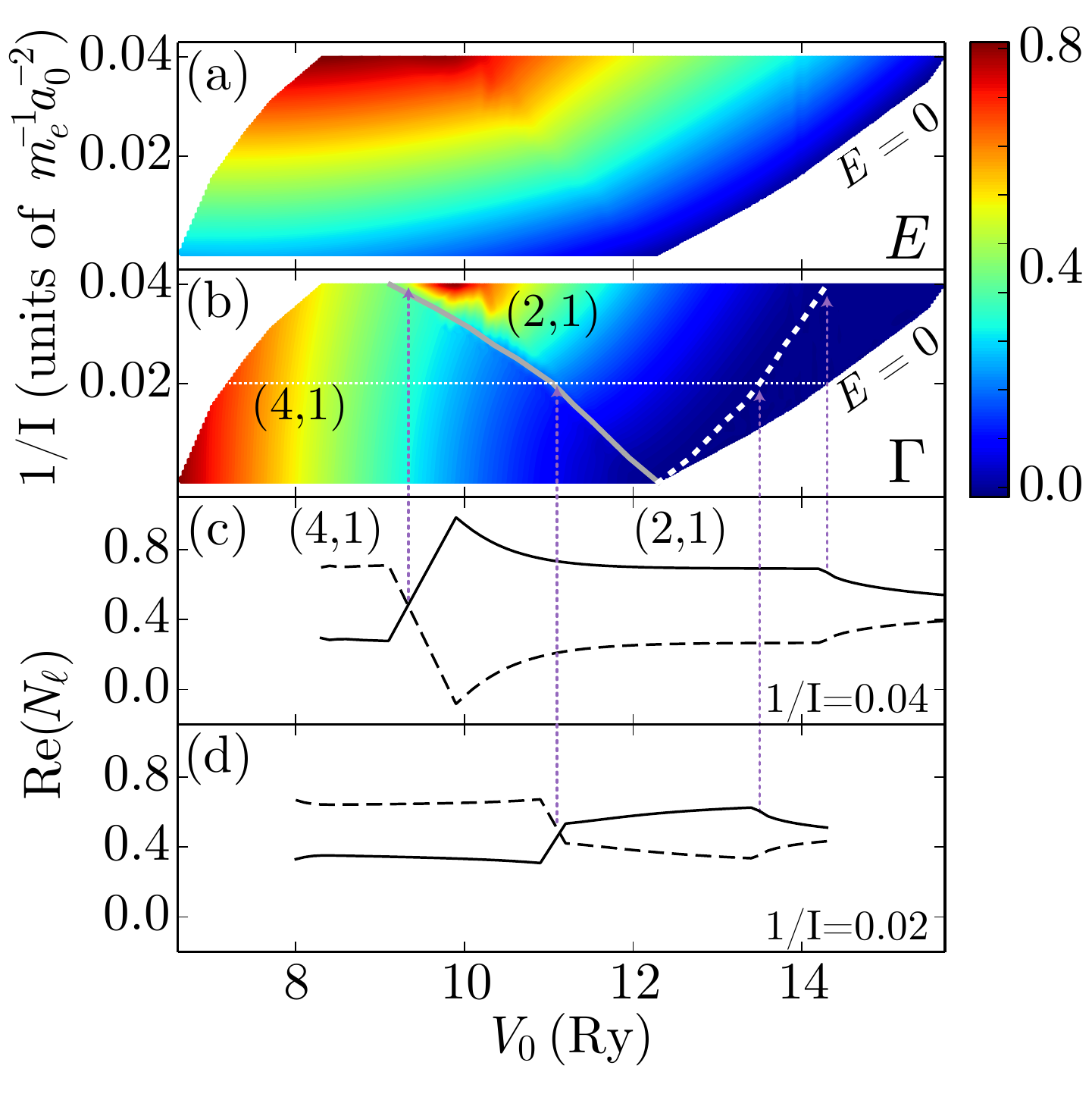}
	\caption{Energy (a) and decay width (b), both in Ry, of the  $3^-_1$ resonance of 
	 the  quadrupolar Gaussian potential with $r_0=a_0$ as a function of the inverse of the moment of inertia and the potential strength. The detachment threshold  ($E=0$) is indicated.
The dominant $(j,\ell)$ channel is marked in panel (b). 
When  the rotational energy of the molecule $E_{\rm rot}^{j=4}$ lies below/above  the energy of  the  $3^-_1$ resonance, the   (4,1) decay channel is open/closed.
The line $E_{\rm rot}^{j=4}=E(3^-_1)$ (thick solid) separating these two regimes is marked, so is the line $E_{\rm rot}^{j=2}=E(3^-_1)$ (thick dotted) which corresponds to the threshold energy for the opening of the (2,1) channel.  The norms of the two dominant channels (2,1) (solid line) and (4,1) (dotted line) are shown as a function of $V_0$ for $1/I=0.04 \,m_e^{-1}a_0^{-2}$ (c)  and 0.02$\,m_e^{-1}a_0^{-2}$  (d).}
	\label{fig:3-_contour}
\end{figure}

We now investigate the impact of the molecular rotation on  the anion's energy spectrum.  
By definition, changing the moment of inertia of the rotor is expected to have a larger effect on states dominated by channels with  large $j$, but in practice such channels  are unlikely to dominate at low energies. 
As an illustrative example, we study the $3_1^-$ state of the  quadrupolar ($\lambda=2$) Gaussian potential. Figure~\ref{fig:3-_contour}(a,b) shows, respectively,  the energy and decay width of the $3_1^-$  resonance as a function of  the potential strength and the inverse moment of inertia.

At large values of $V_0$ when the $3_1^-$ resonance lies close to the threshold, its wave function is primarily described in terms of  two
 channels with  $(j,\ell) = (2,1)$ and $(4,1)$ with the dominant (2,1) amplitude, see Fig.~\ref{fig:3-_contour}(c,d). At a finite value of $I$, as the energy of the resonance increases, a transition takes place to a state dominated by the (4,1) component that is associated with a reduction of the decay width. This transition can be explained in terms of channel coupling. At very low values of $1/I$  the resonance's energy $E(3^-_1)$ lies above the rotational $4^+$ state of the molecule. As the moment of inertia decreases, the $4^+$  member of the ground-state rotational band of the molecule moves up in energy, and at some value of $I$  it becomes degenerate with the energy of the $E(3^-_1)$ resonance, i.e., $ E_{\rm rot}^{j=4}=E(3^-_1)$.
 At still  higher values of $1/I$, the (4,1) channel is closed to the anion's decay. As seen in Fig.~\ref{fig:3-_contour}(b), the irregular behavior seen in the width of the resonance can be attributed to the (4,1) channel closing effect \cite{fossez16_1335}.
A second irregularity in Fig.~\ref{fig:3-_contour}(c,d), seen at  large potential strengths,  corresponds to $E_{\rm rot}^{j=2}=E(3^-_1)$. 
As the resonance approaches the threshold, its tiny decay width can be associated with the  (0,3) channel. Due to its higher centrifugal barrier, (0,3) channel contributes around $1\%$ to the total norm in the threshold region.
		
\subsection{Unbound threshold solutions in the supercritical region}
\label{sub:subspucritical}

\begin{figure}[tb]
	\includegraphics[width=0.9\linewidth]{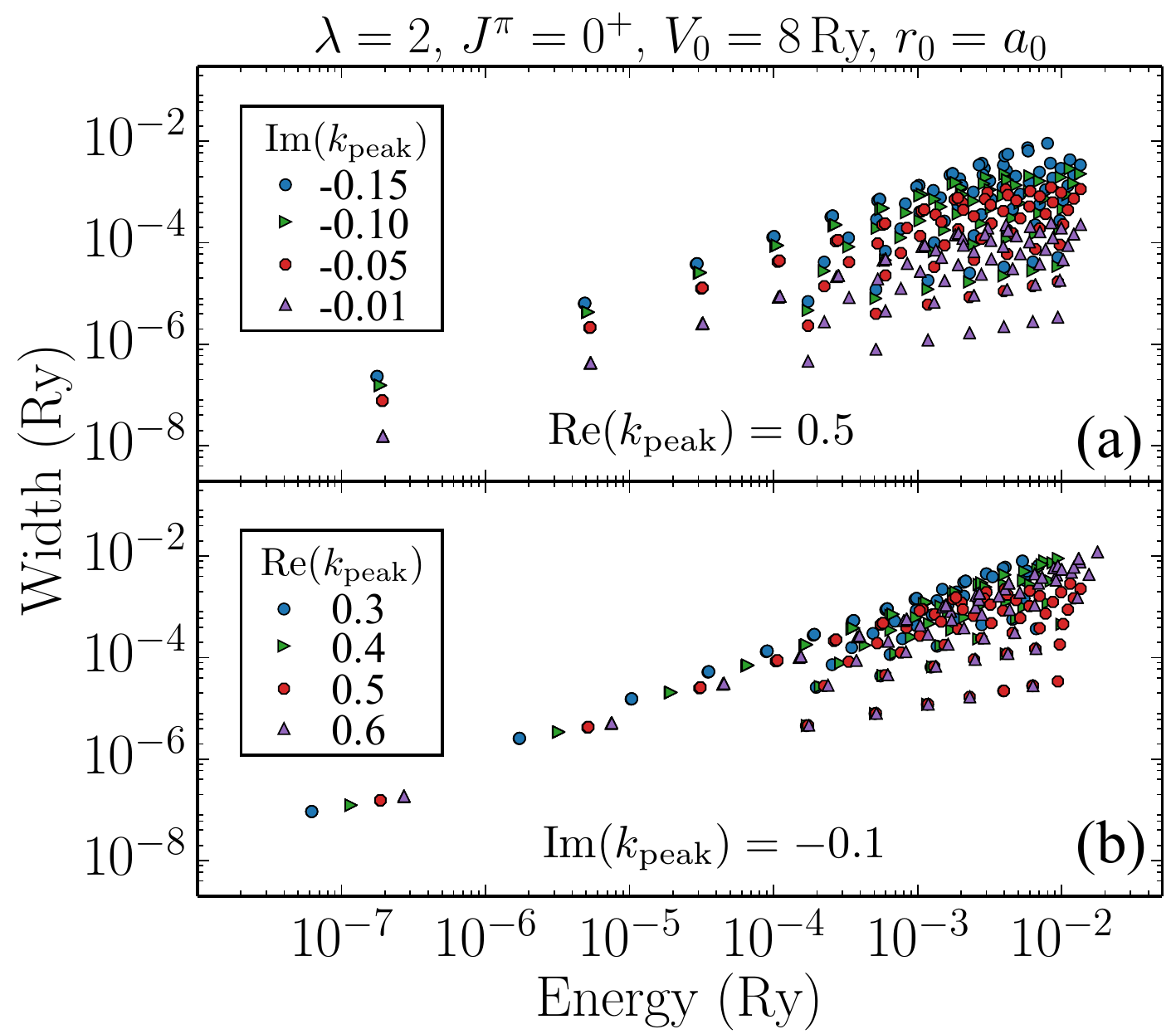}
	\caption{Unbound threshold 0$^+$ states of the quadrupolar Gaussian potential obtained by using scattering ${\cal L}$-contours with different $k_{\rm peak}$ (in units of $a_0^{-1}$) in the complex momentum plane, see Fig.~\ref{figcomplexk}. The potential has $V_0=8$\,Ry and $r_0=a_0$.}
	\label{fig:contour_2_0+}
\end{figure}
In our previous study on quadrupole-bound anions \cite{fossez16_1775}, based on a realistic  pseudopotential, it was shown that there appear series of narrow resonances at energies close to the rotor energies, exhibiting fairly regular patterns. Similar sequences of threshold states, predicted by the present model, are shown in Fig.~\ref{fig:contour_2_0+}, which displays unbound 0$^+$ states of the quadrupolar Gaussian potential computed
 with different scattering contours in the complex momentum plane obtained by varying 
$k_{\rm peak}$, see Fig.~\ref{figcomplexk}. It is seen that the calculated states exhibit appreciable contour dependence.

Similar results have been obtained for other $J^\pi$ states and  Gaussian potentials with higher-multipolarity potentials. Since the  general pattern of near-threshold solutions obtained in different calculations seems to be fairly generic, and primarily depends on the shape of the contour used, they should be interpreted in terms of non-resonant scattering continuum states rather than resonance poles.

\section{Conclusions}\label{s:conclusion}

In this work, we studied  properties of near-threshold states of multipole-bound anions using the Berggren expansion method within the coupled-channel formalism.
We considered a Hamiltonian of a nonadiabatic electron-plus-molecule model with the particle-core interaction being represented by a multipolar Gaussian potential. Such a four-parameter model, rooted in scale-separation arguments of halo effective field theory,  is expected to describe general trends of near-threshold resonant poles for multipolarities $\lambda \ge 2$.

By calculating the threshold lines for anions of different multipolarity, we predicted that within this model, higher-$\lambda$  anions can exist as marginally-bound open systems.
The role of the low-$\ell$ channels in shaping the transition between  subcritical and supercritical regimes  has been explored. We demonstrate the presence of a complex interplay between  bound states,  antibound states, subthreshold resonances, and decaying resonances as the strength of the Gaussian potential is varied. In some cases, we predict the presence of exceptional points. The fact that antibound states and subthreshold resonances can be present in multipolar anions is of interest as they can affect scattering cross sections at low energy. 

For Gaussian potentials, the outgoing boundary condition can be readily imposed. Consequently, the
rotational band of the anion is not affected when the zero-energy threshold is reached. This indicates  that the presence of two coupling regimes  of rotation predicted to exist in realistic calculations for dipole-bound anions~\cite{fossez15_1028} must be due to specific asymptotic behavior of the dipolar pseudo-potential in the presence of molecular rotation.
The non-adiabatic coupling due to the collective rotation of the molecular core
 can give rise to a transition  into the supercritical region. We also predict interesting channel-coupling effects resulting in variation of an anion's decay width due to rotation.

In summary, by looking systematically at the pattern of resonant poles of multipole-bound anions near the electron detachment threshold we uncover a  rich structure of the low-energy continuum. These simple systems  are indeed splendid laboratories  of generic phenomena found in marginally-bound molecules and atomic nuclei.


\begin{acknowledgments}
Discussions with Simin Wang are gratefully acknowledged as well as useful comments from Erik Olsen. The original version of the particle-rotor code used in this work was written by Nicolas Michel.
 This work was supported by the U.S.\ Department of Energy, Office of Science, Office of Nuclear Physics under award number DE-SC0013365 and by the National Science Foundation under award number PHY-1403906.
\end{acknowledgments}


\begin{thebibliography}{82}%
\makeatletter
\providecommand \@ifxundefined [1]{%
 \@ifx{#1\undefined}
}%
\providecommand \@ifnum [1]{%
 \ifnum #1\expandafter \@firstoftwo
 \else \expandafter \@secondoftwo
 \fi
}%
\providecommand \@ifx [1]{%
 \ifx #1\expandafter \@firstoftwo
 \else \expandafter \@secondoftwo
 \fi
}%
\providecommand \natexlab [1]{#1}%
\providecommand \enquote  [1]{``#1''}%
\providecommand \bibnamefont  [1]{#1}%
\providecommand \bibfnamefont [1]{#1}%
\providecommand \citenamefont [1]{#1}%
\providecommand \href@noop [0]{\@secondoftwo}%
\providecommand \href [0]{\begingroup \@sanitize@url \@href}%
\providecommand \@href[1]{\@@startlink{#1}\@@href}%
\providecommand \@@href[1]{\endgroup#1\@@endlink}%
\providecommand \@sanitize@url [0]{\catcode `\\12\catcode `\$12\catcode
  `\&12\catcode `\#12\catcode `\^12\catcode `\_12\catcode `\%12\relax}%
\providecommand \@@startlink[1]{}%
\providecommand \@@endlink[0]{}%
\providecommand \url  [0]{\begingroup\@sanitize@url \@url }%
\providecommand \@url [1]{\endgroup\@href {#1}{\urlprefix }}%
\providecommand \urlprefix  [0]{URL }%
\providecommand \Eprint [0]{\href }%
\providecommand \doibase [0]{http://dx.doi.org/}%
\providecommand \selectlanguage [0]{\@gobble}%
\providecommand \bibinfo  [0]{\@secondoftwo}%
\providecommand \bibfield  [0]{\@secondoftwo}%
\providecommand \translation [1]{[#1]}%
\providecommand \BibitemOpen [0]{}%
\providecommand \bibitemStop [0]{}%
\providecommand \bibitemNoStop [0]{.\EOS\space}%
\providecommand \EOS [0]{\spacefactor3000\relax}%
\providecommand \BibitemShut  [1]{\csname bibitem#1\endcsname}%
\let\auto@bib@innerbib\@empty
\bibitem [{\citenamefont {Desfran\c{c}ois}\ \emph {et~al.}(1996)\citenamefont
  {Desfran\c{c}ois}, \citenamefont {{Abdoul-Carime}},\ and\ \citenamefont
  {Schermann}}]{desfrancois96_114}%
  \BibitemOpen
  \bibfield  {author} {\bibinfo {author} {\bibfnamefont {C.}~\bibnamefont
  {Desfran\c{c}ois}}, \bibinfo {author} {\bibfnamefont {H.}~\bibnamefont
  {{Abdoul-Carime}}}, \ and\ \bibinfo {author} {\bibfnamefont {J.~P.}\
  \bibnamefont {Schermann}},\ }\href
  {https://dx.doi.org/10.1142/S0217979296000520} {\bibfield  {journal}
  {\bibinfo  {journal} {Int. J. Mol. Phys. B}\ }\textbf {\bibinfo {volume}
  {10}},\ \bibinfo {pages} {1339} (\bibinfo {year} {1996})}\BibitemShut
  {NoStop}%
\bibitem [{\citenamefont {Compton}\ and\ \citenamefont
  {Hammer}(2001)}]{compton01_b60}%
  \BibitemOpen
  \bibfield  {author} {\bibinfo {author} {\bibfnamefont {R.~N.}\ \bibnamefont
  {Compton}}\ and\ \bibinfo {author} {\bibfnamefont {N.~I.}\ \bibnamefont
  {Hammer}},\ }\href@noop {} {\emph {\bibinfo {title} {Multipole-{B}ound
  {M}olecular {A}nions}}},\ \bibinfo {edition} {1st}\ ed.\ (\bibinfo
  {publisher} {Elsevier},\ \bibinfo {year} {2001})\BibitemShut {NoStop}%
\bibitem [{\citenamefont {Jordan}\ and\ \citenamefont
  {Wang}(2003)}]{jordan03_352}%
  \BibitemOpen
  \bibfield  {author} {\bibinfo {author} {\bibfnamefont {K.~D.}\ \bibnamefont
  {Jordan}}\ and\ \bibinfo {author} {\bibfnamefont {F.}~\bibnamefont {Wang}},\
  }\href {https://dx.doi.org/10.1146/annurev.physchem.54.011002.103851}
  {\bibfield  {journal} {\bibinfo  {journal} {Annu. Rev. Phys. Chem.}\ }\textbf
  {\bibinfo {volume} {54}},\ \bibinfo {pages} {367} (\bibinfo {year}
  {2003})}\BibitemShut {NoStop}%
\bibitem [{\citenamefont {Simons}(2008)}]{simons08_1079}%
  \BibitemOpen
  \bibfield  {author} {\bibinfo {author} {\bibfnamefont {J.}~\bibnamefont
  {Simons}},\ }\href {https://dx.doi.org/10.1021/jp711490b} {\bibfield
  {journal} {\bibinfo  {journal} {J. Phys. Chem. A}\ }\textbf {\bibinfo
  {volume} {112}},\ \bibinfo {pages} {6401} (\bibinfo {year}
  {2008})}\BibitemShut {NoStop}%
\bibitem [{\citenamefont {Fermi}\ and\ \citenamefont
  {Teller}(1947)}]{fermi47_110}%
  \BibitemOpen
  \bibfield  {author} {\bibinfo {author} {\bibfnamefont {E.}~\bibnamefont
  {Fermi}}\ and\ \bibinfo {author} {\bibfnamefont {E.}~\bibnamefont {Teller}},\
  }\href {https://dx.doi.org/10.1103/PhysRev.72.399} {\bibfield  {journal}
  {\bibinfo  {journal} {Phys. Rev.}\ }\textbf {\bibinfo {volume} {72}},\
  \bibinfo {pages} {399} (\bibinfo {year} {1947})}\BibitemShut {NoStop}%
\bibitem [{\citenamefont {Desfran\c{c}ois}(1995)}]{desfrancois95_205}%
  \BibitemOpen
  \bibfield  {author} {\bibinfo {author} {\bibfnamefont {C.}~\bibnamefont
  {Desfran\c{c}ois}},\ }\href {https://dx.doi.org/10.1103/PhysRevA.51.3667}
  {\bibfield  {journal} {\bibinfo  {journal} {Phys. Rev. A}\ }\textbf {\bibinfo
  {volume} {51}},\ \bibinfo {pages} {3667} (\bibinfo {year}
  {1995})}\BibitemShut {NoStop}%
\bibitem [{\citenamefont {{Abdoul-Carime}}\ and\ \citenamefont
  {Desfran\c{c}ois}(1998)}]{abdoul98_108}%
  \BibitemOpen
  \bibfield  {author} {\bibinfo {author} {\bibfnamefont {H.}~\bibnamefont
  {{Abdoul-Carime}}}\ and\ \bibinfo {author} {\bibfnamefont {C.}~\bibnamefont
  {Desfran\c{c}ois}},\ }\href {https://dx.doi.org/10.1007/s100530050124}
  {\bibfield  {journal} {\bibinfo  {journal} {Eur. Phys. J. D}\ }\textbf
  {\bibinfo {volume} {2}},\ \bibinfo {pages} {149} (\bibinfo {year}
  {1998})}\BibitemShut {NoStop}%
\bibitem [{\citenamefont {{Abdoul-Carime}}\ \emph {et~al.}(2002)\citenamefont
  {{Abdoul-Carime}}, \citenamefont {Schermann},\ and\ \citenamefont
  {Desfran\c{c}ois}}]{abdoul02_1321}%
  \BibitemOpen
  \bibfield  {author} {\bibinfo {author} {\bibfnamefont {H.}~\bibnamefont
  {{Abdoul-Carime}}}, \bibinfo {author} {\bibfnamefont {J.~P.}\ \bibnamefont
  {Schermann}}, \ and\ \bibinfo {author} {\bibfnamefont {C.}~\bibnamefont
  {Desfran\c{c}ois}},\ }\href {https://dx.doi.org/10.1007/s006010200019}
  {\bibfield  {journal} {\bibinfo  {journal} {Few-Body Systems}\ }\textbf
  {\bibinfo {volume} {31}},\ \bibinfo {pages} {183} (\bibinfo {year}
  {2002})}\BibitemShut {NoStop}%
\bibitem [{\citenamefont {Garrett}(1970)}]{garrett70_118}%
  \BibitemOpen
  \bibfield  {author} {\bibinfo {author} {\bibfnamefont {W.~R.}\ \bibnamefont
  {Garrett}},\ }\href {https://dx.doi.org/10.1016/0009-2614(70)80045-8}
  {\bibfield  {journal} {\bibinfo  {journal} {Chem. Phys. Lett.}\ }\textbf
  {\bibinfo {volume} {5}},\ \bibinfo {pages} {393} (\bibinfo {year}
  {1970})}\BibitemShut {NoStop}%
\bibitem [{\citenamefont {Garrett}(1979)}]{garrett71_106}%
  \BibitemOpen
  \bibfield  {author} {\bibinfo {author} {\bibfnamefont {W.~R.}\ \bibnamefont
  {Garrett}},\ }\href {https://dx.doi.org/10.1063/1.438416} {\bibfield
  {journal} {\bibinfo  {journal} {J. Chem. Phys.}\ }\textbf {\bibinfo {volume}
  {71}},\ \bibinfo {pages} {651} (\bibinfo {year} {1979})}\BibitemShut
  {NoStop}%
\bibitem [{\citenamefont {Garrett}(1982)}]{garrett82_104}%
  \BibitemOpen
  \bibfield  {author} {\bibinfo {author} {\bibfnamefont {W.~R.}\ \bibnamefont
  {Garrett}},\ }\href {https://dx.doi.org/10.1063/1.444268} {\bibfield
  {journal} {\bibinfo  {journal} {J. Chem. Phys.}\ }\textbf {\bibinfo {volume}
  {77}},\ \bibinfo {pages} {3666} (\bibinfo {year} {1982})}\BibitemShut
  {NoStop}%
\bibitem [{\citenamefont {Ard}\ \emph {et~al.}(2009)\citenamefont {Ard},
  \citenamefont {Garrett}, \citenamefont {Compton}, \citenamefont {Adamowicz},\
  and\ \citenamefont {Stepanian}}]{ard09_122}%
  \BibitemOpen
  \bibfield  {author} {\bibinfo {author} {\bibfnamefont {S.}~\bibnamefont
  {Ard}}, \bibinfo {author} {\bibfnamefont {W.~R.}\ \bibnamefont {Garrett}},
  \bibinfo {author} {\bibfnamefont {R.~N.}\ \bibnamefont {Compton}}, \bibinfo
  {author} {\bibfnamefont {L.}~\bibnamefont {Adamowicz}}, \ and\ \bibinfo
  {author} {\bibfnamefont {S.~G.}\ \bibnamefont {Stepanian}},\ }\href
  {https://dx.doi.org/10.1016/j.cplett.2009.04.007} {\bibfield  {journal}
  {\bibinfo  {journal} {Chem. Phys. Lett.}\ }\textbf {\bibinfo {volume}
  {473}},\ \bibinfo {pages} {223} (\bibinfo {year} {2009})}\BibitemShut
  {NoStop}%
\bibitem [{\citenamefont {Desfran\c{c}ois}\ \emph {et~al.}(2004)\citenamefont
  {Desfran\c{c}ois}, \citenamefont {Bouteiller}, \citenamefont {Schermann},
  \citenamefont {Radisic}, \citenamefont {Stokes}, \citenamefont {Bowen},
  \citenamefont {Hammer},\ and\ \citenamefont {Compton}}]{desfrancois04_199}%
  \BibitemOpen
  \bibfield  {author} {\bibinfo {author} {\bibfnamefont {C.}~\bibnamefont
  {Desfran\c{c}ois}}, \bibinfo {author} {\bibfnamefont {Y.}~\bibnamefont
  {Bouteiller}}, \bibinfo {author} {\bibfnamefont {J.~P.}\ \bibnamefont
  {Schermann}}, \bibinfo {author} {\bibfnamefont {D.}~\bibnamefont {Radisic}},
  \bibinfo {author} {\bibfnamefont {S.~T.}\ \bibnamefont {Stokes}}, \bibinfo
  {author} {\bibfnamefont {K.~H.}\ \bibnamefont {Bowen}}, \bibinfo {author}
  {\bibfnamefont {N.~I.}\ \bibnamefont {Hammer}}, \ and\ \bibinfo {author}
  {\bibfnamefont {R.~N.}\ \bibnamefont {Compton}},\ }\href
  {https://dx.doi.org/10.1103/PhysRevLett.92.083003} {\bibfield  {journal}
  {\bibinfo  {journal} {Phys. Rev. Lett.}\ }\textbf {\bibinfo {volume} {92}},\
  \bibinfo {pages} {083003} (\bibinfo {year} {2004})}\BibitemShut {NoStop}%
\bibitem [{\citenamefont {Desfran\c{c}ois}\ \emph {et~al.}(1998)\citenamefont
  {Desfran\c{c}ois}, \citenamefont {P\'eriquet}, \citenamefont {Carles},
  \citenamefont {Schermann},\ and\ \citenamefont
  {Adamowicz}}]{desfrancois98_1336}%
  \BibitemOpen
  \bibfield  {author} {\bibinfo {author} {\bibfnamefont {C.}~\bibnamefont
  {Desfran\c{c}ois}}, \bibinfo {author} {\bibfnamefont {V.}~\bibnamefont
  {P\'eriquet}}, \bibinfo {author} {\bibfnamefont {S.}~\bibnamefont {Carles}},
  \bibinfo {author} {\bibfnamefont {J.~P.}\ \bibnamefont {Schermann}}, \ and\
  \bibinfo {author} {\bibfnamefont {L.}~\bibnamefont {Adamowicz}},\ }\href
  {http://dx.doi.org/10.1016/S0301-0104(98)00271-7} {\bibfield  {journal}
  {\bibinfo  {journal} {Chem. Phys.}\ }\textbf {\bibinfo {volume} {239}},\
  \bibinfo {pages} {475} (\bibinfo {year} {1998})}\BibitemShut {NoStop}%
\bibitem [{\citenamefont {Fossez}\ \emph {et~al.}(2013)\citenamefont {Fossez},
  \citenamefont {Michel}, \citenamefont {Nazarewicz},\ and\ \citenamefont
  {P{\l}oszajczak}}]{fossez13_552}%
  \BibitemOpen
  \bibfield  {author} {\bibinfo {author} {\bibfnamefont {K.}~\bibnamefont
  {Fossez}}, \bibinfo {author} {\bibfnamefont {N.}~\bibnamefont {Michel}},
  \bibinfo {author} {\bibfnamefont {W.}~\bibnamefont {Nazarewicz}}, \ and\
  \bibinfo {author} {\bibfnamefont {M.}~\bibnamefont {P{\l}oszajczak}},\ }\href
  {https://dx.doi.org/10.1103/PhysRevA.87.042515} {\bibfield  {journal}
  {\bibinfo  {journal} {Phys. Rev. A}\ }\textbf {\bibinfo {volume} {87}},\
  \bibinfo {pages} {042515} (\bibinfo {year} {2013})}\BibitemShut {NoStop}%
\bibitem [{\citenamefont {Fossez}\ \emph {et~al.}(2015)\citenamefont {Fossez},
  \citenamefont {Michel}, \citenamefont {Nazarewicz}, \citenamefont
  {P{\l}oszajczak},\ and\ \citenamefont {Jaganathen}}]{fossez15_1028}%
  \BibitemOpen
  \bibfield  {author} {\bibinfo {author} {\bibfnamefont {K.}~\bibnamefont
  {Fossez}}, \bibinfo {author} {\bibfnamefont {N.}~\bibnamefont {Michel}},
  \bibinfo {author} {\bibfnamefont {W.}~\bibnamefont {Nazarewicz}}, \bibinfo
  {author} {\bibfnamefont {M.}~\bibnamefont {P{\l}oszajczak}}, \ and\ \bibinfo
  {author} {\bibfnamefont {Y.}~\bibnamefont {Jaganathen}},\ }\href
  {https://dx.doi.org/10.1103/PhysRevA.91.012503} {\bibfield  {journal}
  {\bibinfo  {journal} {Phys. Rev. A}\ }\textbf {\bibinfo {volume} {91}},\
  \bibinfo {pages} {012503} (\bibinfo {year} {2015})}\BibitemShut {NoStop}%
\bibitem [{\citenamefont {Fossez}\ \emph
  {et~al.}(2016{\natexlab{a}})\citenamefont {Fossez}, \citenamefont {Mao},
  \citenamefont {Nazarewicz}, \citenamefont {Michel}, \citenamefont {Garrett},\
  and\ \citenamefont {P{\l}oszajczak}}]{fossez16_1775}%
  \BibitemOpen
  \bibfield  {author} {\bibinfo {author} {\bibfnamefont {K.}~\bibnamefont
  {Fossez}}, \bibinfo {author} {\bibfnamefont {X.}~\bibnamefont {Mao}},
  \bibinfo {author} {\bibfnamefont {W.}~\bibnamefont {Nazarewicz}}, \bibinfo
  {author} {\bibfnamefont {N.}~\bibnamefont {Michel}}, \bibinfo {author}
  {\bibfnamefont {W.~R.}\ \bibnamefont {Garrett}}, \ and\ \bibinfo {author}
  {\bibfnamefont {M.}~\bibnamefont {P{\l}oszajczak}},\ }\href
  {http://dx.doi.org/10.1103/PhysRevA.94.032511} {\bibfield  {journal}
  {\bibinfo  {journal} {Phys. Rev. A}\ }\textbf {\bibinfo {volume} {94}},\
  \bibinfo {pages} {032511} (\bibinfo {year} {2016}{\natexlab{a}})}\BibitemShut
  {NoStop}%
\bibitem [{\citenamefont {Jordan}(1977)}]{jordan77_232}%
  \BibitemOpen
  \bibfield  {author} {\bibinfo {author} {\bibfnamefont {K.~D.}\ \bibnamefont
  {Jordan}},\ }\href {http://link.aip.org/link/doi/10.1063/1.434309} {\bibfield
   {journal} {\bibinfo  {journal} {J. Chem. Phys.}\ }\textbf {\bibinfo {volume}
  {66}},\ \bibinfo {pages} {3305} (\bibinfo {year} {1977})}\BibitemShut
  {NoStop}%
\bibitem [{\citenamefont {Gutsev}\ \emph
  {et~al.}(1998{\natexlab{a}})\citenamefont {Gutsev}, \citenamefont {Nooijen},\
  and\ \citenamefont {Bartlett}}]{gutsev98_1097}%
  \BibitemOpen
  \bibfield  {author} {\bibinfo {author} {\bibfnamefont {G.~L.}\ \bibnamefont
  {Gutsev}}, \bibinfo {author} {\bibfnamefont {M.}~\bibnamefont {Nooijen}}, \
  and\ \bibinfo {author} {\bibfnamefont {R.~J.}\ \bibnamefont {Bartlett}},\
  }\href {https://dx.doi.org/10.1103/PhysRevA.57.1646} {\bibfield  {journal}
  {\bibinfo  {journal} {Phys. Rev. A}\ }\textbf {\bibinfo {volume} {57}},\
  \bibinfo {pages} {1646} (\bibinfo {year} {1998}{\natexlab{a}})}\BibitemShut
  {NoStop}%
\bibitem [{\citenamefont {Adamowicz}(1989)}]{adamowicz89_346}%
  \BibitemOpen
  \bibfield  {author} {\bibinfo {author} {\bibfnamefont {L.}~\bibnamefont
  {Adamowicz}},\ }\href {http://link.aip.org/link/doi/10.1063/1.457246}
  {\bibfield  {journal} {\bibinfo  {journal} {J. Chem. Phys.}\ }\textbf
  {\bibinfo {volume} {91}},\ \bibinfo {pages} {7787} (\bibinfo {year}
  {1989})}\BibitemShut {NoStop}%
\bibitem [{\citenamefont {Smith}\ \emph {et~al.}(1999)\citenamefont {Smith},
  \citenamefont {Smets},\ and\ \citenamefont {Adamowicz}}]{smith99_355}%
  \BibitemOpen
  \bibfield  {author} {\bibinfo {author} {\bibfnamefont {D.~M.~A.}\
  \bibnamefont {Smith}}, \bibinfo {author} {\bibfnamefont {J.}~\bibnamefont
  {Smets}}, \ and\ \bibinfo {author} {\bibfnamefont {L.}~\bibnamefont
  {Adamowicz}},\ }\href {http://link.aip.org/link/doi/10.1063/1.478796}
  {\bibfield  {journal} {\bibinfo  {journal} {J. Chem. Phys.}\ }\textbf
  {\bibinfo {volume} {110}},\ \bibinfo {pages} {3804} (\bibinfo {year}
  {1999})}\BibitemShut {NoStop}%
\bibitem [{\citenamefont {Clary}\ and\ \citenamefont
  {Benoit}(1999)}]{clary99_356}%
  \BibitemOpen
  \bibfield  {author} {\bibinfo {author} {\bibfnamefont {D.~C.}\ \bibnamefont
  {Clary}}\ and\ \bibinfo {author} {\bibfnamefont {D.~M.}\ \bibnamefont
  {Benoit}},\ }\href {http://link.aip.org/link/doi/10.1063/1.480409} {\bibfield
   {journal} {\bibinfo  {journal} {J. Chem. Phys.}\ }\textbf {\bibinfo {volume}
  {111}},\ \bibinfo {pages} {10559} (\bibinfo {year} {1999})}\BibitemShut
  {NoStop}%
\bibitem [{\citenamefont {Kalcher}\ and\ \citenamefont
  {Sax}(2000)}]{kalcher00_560}%
  \BibitemOpen
  \bibfield  {author} {\bibinfo {author} {\bibfnamefont {J.}~\bibnamefont
  {Kalcher}}\ and\ \bibinfo {author} {\bibfnamefont {A.~F.}\ \bibnamefont
  {Sax}},\ }\href {https://dx.doi.org/10.1016/S0009-2614(00)00754-5} {\bibfield
   {journal} {\bibinfo  {journal} {Chem. Phys. Lett.}\ }\textbf {\bibinfo
  {volume} {326}},\ \bibinfo {pages} {80} (\bibinfo {year} {2000})}\BibitemShut
  {NoStop}%
\bibitem [{\citenamefont {Skurski}\ \emph {et~al.}(2002)\citenamefont
  {Skurski}, \citenamefont {D\c{a}bkowska}, \citenamefont {Sawicka},\ and\
  \citenamefont {Rak}}]{skurski02_575}%
  \BibitemOpen
  \bibfield  {author} {\bibinfo {author} {\bibfnamefont {P.}~\bibnamefont
  {Skurski}}, \bibinfo {author} {\bibfnamefont {I.}~\bibnamefont
  {D\c{a}bkowska}}, \bibinfo {author} {\bibfnamefont {A.}~\bibnamefont
  {Sawicka}}, \ and\ \bibinfo {author} {\bibfnamefont {J.}~\bibnamefont
  {Rak}},\ }\href {https://dx.doi.org/10.1016/S0301-0104(02)00488-3} {\bibfield
   {journal} {\bibinfo  {journal} {Chem. Phys.}\ }\textbf {\bibinfo {volume}
  {279}},\ \bibinfo {pages} {101} (\bibinfo {year} {2002})}\BibitemShut
  {NoStop}%
\bibitem [{\citenamefont {Peterson}\ and\ \citenamefont
  {Gutowski}(2002)}]{peterson02_123}%
  \BibitemOpen
  \bibfield  {author} {\bibinfo {author} {\bibfnamefont {K.~A.}\ \bibnamefont
  {Peterson}}\ and\ \bibinfo {author} {\bibfnamefont {M.}~\bibnamefont
  {Gutowski}},\ }\href {https://dx.doi.org/10.1063/1.1445743} {\bibfield
  {journal} {\bibinfo  {journal} {J. Chem. Phys.}\ }\textbf {\bibinfo {volume}
  {116}},\ \bibinfo {pages} {3297} (\bibinfo {year} {2002})}\BibitemShut
  {NoStop}%
\bibitem [{\citenamefont {Sommerfeld}(2004)}]{sommerfeld04_1573}%
  \BibitemOpen
  \bibfield  {author} {\bibinfo {author} {\bibfnamefont {T.}~\bibnamefont
  {Sommerfeld}},\ }\href {http://dx.doi.org/10.1063/1.1774979} {\bibfield
  {journal} {\bibinfo  {journal} {J. Chem. Phys.}\ }\textbf {\bibinfo {volume}
  {121}},\ \bibinfo {pages} {4097} (\bibinfo {year} {2004})}\BibitemShut
  {NoStop}%
\bibitem [{\citenamefont {Sommerfeld}\ \emph {et~al.}(2014)\citenamefont
  {Sommerfeld}, \citenamefont {Dreux},\ and\ \citenamefont
  {Joshi}}]{sommerfeld14_1187}%
  \BibitemOpen
  \bibfield  {author} {\bibinfo {author} {\bibfnamefont {T.}~\bibnamefont
  {Sommerfeld}}, \bibinfo {author} {\bibfnamefont {K.~M.}\ \bibnamefont
  {Dreux}}, \ and\ \bibinfo {author} {\bibfnamefont {R.}~\bibnamefont
  {Joshi}},\ }\href {dx.doi.org/10.1021/jp411787w} {\bibfield  {journal}
  {\bibinfo  {journal} {J. Phys. Chem. A}\ }\textbf {\bibinfo {volume} {118}},\
  \bibinfo {pages} {7320} (\bibinfo {year} {2014})}\BibitemShut {NoStop}%
\bibitem [{\citenamefont {Gutsev}\ and\ \citenamefont
  {Jena}(1999)}]{gutsev99_1324}%
  \BibitemOpen
  \bibfield  {author} {\bibinfo {author} {\bibfnamefont {G.~L.}\ \bibnamefont
  {Gutsev}}\ and\ \bibinfo {author} {\bibfnamefont {P.}~\bibnamefont {Jena}},\
  }\href {http://dx.doi.org/10.1063/1.480262} {\bibfield  {journal} {\bibinfo
  {journal} {J. Chem. Phys.}\ }\textbf {\bibinfo {volume} {111}},\ \bibinfo
  {pages} {504} (\bibinfo {year} {1999})}\BibitemShut {NoStop}%
\bibitem [{\citenamefont {Gutsev}\ \emph
  {et~al.}(1998{\natexlab{b}})\citenamefont {Gutsev}, \citenamefont
  {Bartlett},\ and\ \citenamefont {Compton}}]{gutsev98_1317}%
  \BibitemOpen
  \bibfield  {author} {\bibinfo {author} {\bibfnamefont {G.~L.}\ \bibnamefont
  {Gutsev}}, \bibinfo {author} {\bibfnamefont {R.~J.}\ \bibnamefont
  {Bartlett}}, \ and\ \bibinfo {author} {\bibfnamefont {R.~N.}\ \bibnamefont
  {Compton}},\ }\href {http://dx.doi.org/10.1063/1.476091} {\bibfield
  {journal} {\bibinfo  {journal} {J. Chem. Phys.}\ }\textbf {\bibinfo {volume}
  {108}},\ \bibinfo {pages} {6756} (\bibinfo {year}
  {1998}{\natexlab{b}})}\BibitemShut {NoStop}%
\bibitem [{\citenamefont {Gutowski}\ and\ \citenamefont
  {Skurski}(1999)}]{gutowski99_1186}%
  \BibitemOpen
  \bibfield  {author} {\bibinfo {author} {\bibfnamefont {M.}~\bibnamefont
  {Gutowski}}\ and\ \bibinfo {author} {\bibfnamefont {P.}~\bibnamefont
  {Skurski}},\ }\href {https://dx.doi.org/10.1016/S0009-2614(99)00172-4}
  {\bibfield  {journal} {\bibinfo  {journal} {Chem. Phys. Lett.}\ }\textbf
  {\bibinfo {volume} {303}},\ \bibinfo {pages} {65} (\bibinfo {year}
  {1999})}\BibitemShut {NoStop}%
\bibitem [{\citenamefont {Herrick}\ and\ \citenamefont
  {Engelking}(1984)}]{herrick84_1248}%
  \BibitemOpen
  \bibfield  {author} {\bibinfo {author} {\bibfnamefont {D.~R.}\ \bibnamefont
  {Herrick}}\ and\ \bibinfo {author} {\bibfnamefont {P.~C.}\ \bibnamefont
  {Engelking}},\ }\href {https://dx.doi.org/10.1103/PhysRevA.29.2421}
  {\bibfield  {journal} {\bibinfo  {journal} {Phys. Rev. A}\ }\textbf {\bibinfo
  {volume} {29}},\ \bibinfo {pages} {2421} (\bibinfo {year}
  {1984})}\BibitemShut {NoStop}%
\bibitem [{\citenamefont {Clary}(1988)}]{clary88_345}%
  \BibitemOpen
  \bibfield  {author} {\bibinfo {author} {\bibfnamefont {D.~C.}\ \bibnamefont
  {Clary}},\ }\href {https://dx.doi.org/10.1021/j100322a028} {\bibfield
  {journal} {\bibinfo  {journal} {J. Phys. Chem.}\ }\textbf {\bibinfo {volume}
  {92}},\ \bibinfo {pages} {3173} (\bibinfo {year} {1988})}\BibitemShut
  {NoStop}%
\bibitem [{\citenamefont {Clary}(1989)}]{clary89_303}%
  \BibitemOpen
  \bibfield  {author} {\bibinfo {author} {\bibfnamefont {D.~C.}\ \bibnamefont
  {Clary}},\ }\href {https://dx.doi.org/10.1103/PhysRevA.40.4392} {\bibfield
  {journal} {\bibinfo  {journal} {Phys. Rev. A}\ }\textbf {\bibinfo {volume}
  {40}},\ \bibinfo {pages} {4392} (\bibinfo {year} {1989})}\BibitemShut
  {NoStop}%
\bibitem [{\citenamefont {Brinkman}\ \emph {et~al.}(1993)\citenamefont
  {Brinkman}, \citenamefont {Berger}, \citenamefont {Marks},\ and\
  \citenamefont {Brauman}}]{brinkman93_208}%
  \BibitemOpen
  \bibfield  {author} {\bibinfo {author} {\bibfnamefont {E.~A.}\ \bibnamefont
  {Brinkman}}, \bibinfo {author} {\bibfnamefont {S.}~\bibnamefont {Berger}},
  \bibinfo {author} {\bibfnamefont {J.}~\bibnamefont {Marks}}, \ and\ \bibinfo
  {author} {\bibfnamefont {J.~I.}\ \bibnamefont {Brauman}},\ }\href
  {http://link.aip.org/link/doi/10.1063/1.465688} {\bibfield  {journal}
  {\bibinfo  {journal} {J. Chem. Phys.}\ }\textbf {\bibinfo {volume} {99}},\
  \bibinfo {pages} {7586} (\bibinfo {year} {1993})}\BibitemShut {NoStop}%
\bibitem [{\citenamefont {Garrett}(2010)}]{garrett10_117}%
  \BibitemOpen
  \bibfield  {author} {\bibinfo {author} {\bibfnamefont {W.~R.}\ \bibnamefont
  {Garrett}},\ }\href {https://dx.doi.org/10.1063/1.3511638} {\bibfield
  {journal} {\bibinfo  {journal} {J. Chem. Phys.}\ }\textbf {\bibinfo {volume}
  {133}},\ \bibinfo {pages} {224103} (\bibinfo {year} {2010})}\BibitemShut
  {NoStop}%
\bibitem [{\citenamefont {Klahn}\ and\ \citenamefont
  {Krebs}(1998)}]{klahn98_752}%
  \BibitemOpen
  \bibfield  {author} {\bibinfo {author} {\bibfnamefont {T.}~\bibnamefont
  {Klahn}}\ and\ \bibinfo {author} {\bibfnamefont {P.}~\bibnamefont {Krebs}},\
  }\href {https://dx.doi.org/10.1063/1.476969} {\bibfield  {journal} {\bibinfo
  {journal} {J. Chem. Phys.}\ }\textbf {\bibinfo {volume} {109}},\ \bibinfo
  {pages} {531} (\bibinfo {year} {1998})}\BibitemShut {NoStop}%
\bibitem [{\citenamefont {{von Stecher}}\ \emph {et~al.}(2009)\citenamefont
  {{von Stecher}}, \citenamefont {D'Incao},\ and\ \citenamefont
  {Greene}}]{Stecher2009}%
  \BibitemOpen
  \bibfield  {author} {\bibinfo {author} {\bibfnamefont {J.}~\bibnamefont {{von
  Stecher}}}, \bibinfo {author} {\bibfnamefont {J.~P.}\ \bibnamefont
  {D'Incao}}, \ and\ \bibinfo {author} {\bibfnamefont {C.~H.}\ \bibnamefont
  {Greene}},\ }\href {\doibase 10.1038/nphys1253} {\bibfield  {journal}
  {\bibinfo  {journal} {Nature Phys.}\ }\textbf {\bibinfo {volume} {5}},\
  \bibinfo {pages} {417 EP} (\bibinfo {year} {2009})}\BibitemShut {NoStop}%
\bibitem [{\citenamefont {Hadizadeh}\ \emph {et~al.}(2011)\citenamefont
  {Hadizadeh}, \citenamefont {Yamashita}, \citenamefont {Tomio}, \citenamefont
  {Delfino},\ and\ \citenamefont {Frederico}}]{Hadizadeh2011}%
  \BibitemOpen
  \bibfield  {author} {\bibinfo {author} {\bibfnamefont {M.~R.}\ \bibnamefont
  {Hadizadeh}}, \bibinfo {author} {\bibfnamefont {M.~T.}\ \bibnamefont
  {Yamashita}}, \bibinfo {author} {\bibfnamefont {L.}~\bibnamefont {Tomio}},
  \bibinfo {author} {\bibfnamefont {A.}~\bibnamefont {Delfino}}, \ and\
  \bibinfo {author} {\bibfnamefont {T.}~\bibnamefont {Frederico}},\ }\href
  {\doibase 10.1103/PhysRevLett.107.135304} {\bibfield  {journal} {\bibinfo
  {journal} {Phys. Rev. Lett.}\ }\textbf {\bibinfo {volume} {107}},\ \bibinfo
  {pages} {135304} (\bibinfo {year} {2011})}\BibitemShut {NoStop}%
\bibitem [{\citenamefont {Hadizadeh}\ \emph {et~al.}(2012)\citenamefont
  {Hadizadeh}, \citenamefont {Yamashita}, \citenamefont {Tomio}, \citenamefont
  {Delfino},\ and\ \citenamefont {Frederico}}]{Hadizadeh2012}%
  \BibitemOpen
  \bibfield  {author} {\bibinfo {author} {\bibfnamefont {M.~R.}\ \bibnamefont
  {Hadizadeh}}, \bibinfo {author} {\bibfnamefont {M.~T.}\ \bibnamefont
  {Yamashita}}, \bibinfo {author} {\bibfnamefont {L.}~\bibnamefont {Tomio}},
  \bibinfo {author} {\bibfnamefont {A.}~\bibnamefont {Delfino}}, \ and\
  \bibinfo {author} {\bibfnamefont {T.}~\bibnamefont {Frederico}},\ }\href
  {\doibase 10.1063/1.3688794} {\bibfield  {journal} {\bibinfo  {journal} {AIP
  Conf. Proc.}\ }\textbf {\bibinfo {volume} {1423}},\ \bibinfo {pages} {130}
  (\bibinfo {year} {2012})}\BibitemShut {NoStop}%
\bibitem [{\citenamefont {Lazauskas}\ and\ \citenamefont
  {Carbonell}(2013)}]{Lazauskas2013}%
  \BibitemOpen
  \bibfield  {author} {\bibinfo {author} {\bibfnamefont {R.}~\bibnamefont
  {Lazauskas}}\ and\ \bibinfo {author} {\bibfnamefont {J.}~\bibnamefont
  {Carbonell}},\ }\href {\doibase 10.1007/s00601-012-0531-y} {\bibfield
  {journal} {\bibinfo  {journal} {Few-Body Systems}\ }\textbf {\bibinfo
  {volume} {54}},\ \bibinfo {pages} {967} (\bibinfo {year} {2013})}\BibitemShut
  {NoStop}%
\bibitem [{\citenamefont {Kievsky}\ \emph {et~al.}(2014)\citenamefont
  {Kievsky}, \citenamefont {Gattobigio},\ and\ \citenamefont
  {Garrido}}]{Kievsky2014}%
  \BibitemOpen
  \bibfield  {author} {\bibinfo {author} {\bibfnamefont {A.}~\bibnamefont
  {Kievsky}}, \bibinfo {author} {\bibfnamefont {M.}~\bibnamefont {Gattobigio}},
  \ and\ \bibinfo {author} {\bibfnamefont {E.}~\bibnamefont {Garrido}},\ }\href
  {http://stacks.iop.org/1742-6596/527/i=1/a=012001} {\bibfield  {journal}
  {\bibinfo  {journal} {J. Phys. Conf. Ser.}\ }\textbf {\bibinfo {volume}
  {527}},\ \bibinfo {pages} {012001} (\bibinfo {year} {2014})}\BibitemShut
  {NoStop}%
\bibitem [{\citenamefont {K\"onig}\ \emph {et~al.}(2017)\citenamefont
  {K\"onig}, \citenamefont {Grie{\ss}hammer}, \citenamefont {Hammer},\ and\
  \citenamefont {{van Kolck}}}]{konig17_1986}%
  \BibitemOpen
  \bibfield  {author} {\bibinfo {author} {\bibfnamefont {S.}~\bibnamefont
  {K\"onig}}, \bibinfo {author} {\bibfnamefont {H.~W.}\ \bibnamefont
  {Grie{\ss}hammer}}, \bibinfo {author} {\bibfnamefont {H.~W.}\ \bibnamefont
  {Hammer}}, \ and\ \bibinfo {author} {\bibfnamefont {U.}~\bibnamefont {{van
  Kolck}}},\ }\href {https://doi.org/10.1103/PhysRevLett.118.202501} {\bibfield
   {journal} {\bibinfo  {journal} {Phys. Rev. Lett.}\ }\textbf {\bibinfo
  {volume} {118}},\ \bibinfo {pages} {202501} (\bibinfo {year}
  {2017})}\BibitemShut {NoStop}%
\bibitem [{\citenamefont {\'Alvarez-Rodr\'{\i}guez}\ \emph
  {et~al.}(2016)\citenamefont {\'Alvarez-Rodr\'{\i}guez}, \citenamefont
  {Deltuva}, \citenamefont {Gattobigio},\ and\ \citenamefont
  {Kievsky}}]{Deltuva2016}%
  \BibitemOpen
  \bibfield  {author} {\bibinfo {author} {\bibfnamefont {R.}~\bibnamefont
  {\'Alvarez-Rodr\'{\i}guez}}, \bibinfo {author} {\bibfnamefont
  {A.}~\bibnamefont {Deltuva}}, \bibinfo {author} {\bibfnamefont
  {M.}~\bibnamefont {Gattobigio}}, \ and\ \bibinfo {author} {\bibfnamefont
  {A.}~\bibnamefont {Kievsky}},\ }\href {\doibase 10.1103/PhysRevA.93.062701}
  {\bibfield  {journal} {\bibinfo  {journal} {Phys. Rev. A}\ }\textbf {\bibinfo
  {volume} {93}},\ \bibinfo {pages} {062701} (\bibinfo {year}
  {2016})}\BibitemShut {NoStop}%
\bibitem [{\citenamefont {Deltuva}(2017)}]{Deltuva2017}%
  \BibitemOpen
  \bibfield  {author} {\bibinfo {author} {\bibfnamefont {A.}~\bibnamefont
  {Deltuva}},\ }\href {\doibase 10.1103/PhysRevA.96.022701} {\bibfield
  {journal} {\bibinfo  {journal} {Phys. Rev. A}\ }\textbf {\bibinfo {volume}
  {96}},\ \bibinfo {pages} {022701} (\bibinfo {year} {2017})}\BibitemShut
  {NoStop}%
\bibitem [{\citenamefont {Shalchi}\ \emph {et~al.}(2017)\citenamefont
  {Shalchi}, \citenamefont {Yamashita}, \citenamefont {Hadizadeh},
  \citenamefont {Frederico},\ and\ \citenamefont {Tomio}}]{Shalchi2017}%
  \BibitemOpen
  \bibfield  {author} {\bibinfo {author} {\bibfnamefont {M.}~\bibnamefont
  {Shalchi}}, \bibinfo {author} {\bibfnamefont {M.}~\bibnamefont {Yamashita}},
  \bibinfo {author} {\bibfnamefont {M.}~\bibnamefont {Hadizadeh}}, \bibinfo
  {author} {\bibfnamefont {T.}~\bibnamefont {Frederico}}, \ and\ \bibinfo
  {author} {\bibfnamefont {L.}~\bibnamefont {Tomio}},\ }\href {\doibase
  10.1016/j.physletb.2016.11.030} {\bibfield  {journal} {\bibinfo  {journal}
  {Phys. Lett. B}\ }\textbf {\bibinfo {volume} {764}},\ \bibinfo {pages} {196 }
  (\bibinfo {year} {2017})}\BibitemShut {NoStop}%
\bibitem [{\citenamefont {Miller}(2018)}]{Miller2018}%
  \BibitemOpen
  \bibfield  {author} {\bibinfo {author} {\bibfnamefont {G.~A.}\ \bibnamefont
  {Miller}},\ }\href {\doibase 10.1016/j.physletb.2017.12.063} {\bibfield
  {journal} {\bibinfo  {journal} {Phys. Lett. B}\ }\textbf {\bibinfo {volume}
  {777}},\ \bibinfo {pages} {442} (\bibinfo {year} {2018})}\BibitemShut
  {NoStop}%
\bibitem [{\citenamefont {Braaten}\ and\ \citenamefont
  {Hammer}(2006)}]{braaten06_823}%
  \BibitemOpen
  \bibfield  {author} {\bibinfo {author} {\bibfnamefont {E.}~\bibnamefont
  {Braaten}}\ and\ \bibinfo {author} {\bibfnamefont {H.~W.}\ \bibnamefont
  {Hammer}},\ }\href {https://dx.doi.org/10.1016/j.physrep.2006.03.001}
  {\bibfield  {journal} {\bibinfo  {journal} {Phys. Rep.}\ }\textbf {\bibinfo
  {volume} {428}},\ \bibinfo {pages} {259} (\bibinfo {year}
  {2006})}\BibitemShut {NoStop}%
\bibitem [{\citenamefont {Bertulani}\ \emph {et~al.}(2002)\citenamefont
  {Bertulani}, \citenamefont {Hammer},\ and\ \citenamefont {{van
  Kolck}}}]{bertulani02_869}%
  \BibitemOpen
  \bibfield  {author} {\bibinfo {author} {\bibfnamefont {C.~A.}\ \bibnamefont
  {Bertulani}}, \bibinfo {author} {\bibfnamefont {H.~W.}\ \bibnamefont
  {Hammer}}, \ and\ \bibinfo {author} {\bibfnamefont {U.}~\bibnamefont {{van
  Kolck}}},\ }\href {https://dx.doi.org/10.1016/S0375-9474(02)01270-8}
  {\bibfield  {journal} {\bibinfo  {journal} {Nucl. Phys. A}\ }\textbf
  {\bibinfo {volume} {712}},\ \bibinfo {pages} {37} (\bibinfo {year}
  {2002})}\BibitemShut {NoStop}%
\bibitem [{\citenamefont {Bedaque}\ \emph {et~al.}(2003)\citenamefont
  {Bedaque}, \citenamefont {Hammer},\ and\ \citenamefont {{van
  Kolck}}}]{bedaque03_1085}%
  \BibitemOpen
  \bibfield  {author} {\bibinfo {author} {\bibfnamefont {P.~F.}\ \bibnamefont
  {Bedaque}}, \bibinfo {author} {\bibfnamefont {H.~W.}\ \bibnamefont {Hammer}},
  \ and\ \bibinfo {author} {\bibfnamefont {U.}~\bibnamefont {{van Kolck}}},\
  }\href {https://dx.doi.org/10.1016/j.physletb.2003.07.049} {\bibfield
  {journal} {\bibinfo  {journal} {Phys. Lett. B}\ }\textbf {\bibinfo {volume}
  {569}},\ \bibinfo {pages} {159} (\bibinfo {year} {2003})}\BibitemShut
  {NoStop}%
\bibitem [{\citenamefont {Hammer}\ and\ \citenamefont
  {Platter}(2010)}]{hammer10_1093}%
  \BibitemOpen
  \bibfield  {author} {\bibinfo {author} {\bibfnamefont {H.~W.}\ \bibnamefont
  {Hammer}}\ and\ \bibinfo {author} {\bibfnamefont {L.}~\bibnamefont
  {Platter}},\ }\href {https://dx.doi.org/10.1146/annurev.nucl.012809.104439}
  {\bibfield  {journal} {\bibinfo  {journal} {Annu. Rev. Nucl. Part. Sci.}\
  }\textbf {\bibinfo {volume} {60}},\ \bibinfo {pages} {207} (\bibinfo {year}
  {2010})}\BibitemShut {NoStop}%
\bibitem [{\citenamefont {Hammer}\ and\ \citenamefont
  {Furnstahl}(2000)}]{hammer00_1683}%
  \BibitemOpen
  \bibfield  {author} {\bibinfo {author} {\bibfnamefont {H.~W.}\ \bibnamefont
  {Hammer}}\ and\ \bibinfo {author} {\bibfnamefont {R.~J.}\ \bibnamefont
  {Furnstahl}},\ }\href {http://dx.doi.org/10.1016/S0375-9474(00)00325-0}
  {\bibfield  {journal} {\bibinfo  {journal} {Nucl. Phys. A}\ }\textbf
  {\bibinfo {volume} {678}},\ \bibinfo {pages} {277} (\bibinfo {year}
  {2000})}\BibitemShut {NoStop}%
\bibitem [{\citenamefont {Hammer}\ \emph {et~al.}(2017)\citenamefont {Hammer},
  \citenamefont {Ji},\ and\ \citenamefont {Phillips}}]{hammer17_1959}%
  \BibitemOpen
  \bibfield  {author} {\bibinfo {author} {\bibfnamefont {H.~W.}\ \bibnamefont
  {Hammer}}, \bibinfo {author} {\bibfnamefont {C.}~\bibnamefont {Ji}}, \ and\
  \bibinfo {author} {\bibfnamefont {D.~R.}\ \bibnamefont {Phillips}},\ }\href
  {https://doi.org/10.1088/1361-6471/aa83db} {\bibfield  {journal} {\bibinfo
  {journal} {J. Phys. G}\ }\textbf {\bibinfo {volume} {44}},\ \bibinfo {pages}
  {103002} (\bibinfo {year} {2017})}\BibitemShut {NoStop}%
\bibitem [{\citenamefont {Herzberg}(1950)}]{Herzberg1}%
  \BibitemOpen
  \bibfield  {author} {\bibinfo {author} {\bibfnamefont {G.}~\bibnamefont
  {Herzberg}},\ }\href@noop {} {\emph {\bibinfo {title} {Molecular spectra and
  molecular structure. Vol.1: Spectra of diatomic molecules}}}\ (\bibinfo
  {publisher} {Van Nostrand Reinhold, New York},\ \bibinfo {year}
  {1950})\BibitemShut {NoStop}%
\bibitem [{\citenamefont {Hagen}\ and\ \citenamefont
  {Vaagen}(2006)}]{hagen06_464}%
  \BibitemOpen
  \bibfield  {author} {\bibinfo {author} {\bibfnamefont {G.}~\bibnamefont
  {Hagen}}\ and\ \bibinfo {author} {\bibfnamefont {J.~S.}\ \bibnamefont
  {Vaagen}},\ }\href {https://dx.doi.org/10.1103/PhysRevC.73.034321} {\bibfield
   {journal} {\bibinfo  {journal} {Phys. Rev. C}\ }\textbf {\bibinfo {volume}
  {73}},\ \bibinfo {pages} {034321} (\bibinfo {year} {2006})}\BibitemShut
  {NoStop}%
\bibitem [{\citenamefont {Papadimitriou}\ \emph {et~al.}(2011)\citenamefont
  {Papadimitriou}, \citenamefont {Kruppa}, \citenamefont {Michel},
  \citenamefont {Nazarewicz}, \citenamefont {P{\l}oszajczak},\ and\
  \citenamefont {Rotureau}}]{papadimitriou11_277}%
  \BibitemOpen
  \bibfield  {author} {\bibinfo {author} {\bibfnamefont {G.}~\bibnamefont
  {Papadimitriou}}, \bibinfo {author} {\bibfnamefont {A.~T.}\ \bibnamefont
  {Kruppa}}, \bibinfo {author} {\bibfnamefont {N.}~\bibnamefont {Michel}},
  \bibinfo {author} {\bibfnamefont {W.}~\bibnamefont {Nazarewicz}}, \bibinfo
  {author} {\bibfnamefont {M.}~\bibnamefont {P{\l}oszajczak}}, \ and\ \bibinfo
  {author} {\bibfnamefont {J.}~\bibnamefont {Rotureau}},\ }\href
  {https://dx.doi.org/10.1103/PhysRevC.84.051304} {\bibfield  {journal}
  {\bibinfo  {journal} {Phys. Rev. C}\ }\textbf {\bibinfo {volume} {84}},\
  \bibinfo {pages} {051304(R)} (\bibinfo {year} {2011})}\BibitemShut {NoStop}%
\bibitem [{\citenamefont {Fossez}\ \emph
  {et~al.}(2016{\natexlab{b}})\citenamefont {Fossez}, \citenamefont
  {Nazarewicz}, \citenamefont {Jaganathen}, \citenamefont {Michel},\ and\
  \citenamefont {P{\l}oszajczak}}]{fossez16_1335}%
  \BibitemOpen
  \bibfield  {author} {\bibinfo {author} {\bibfnamefont {K.}~\bibnamefont
  {Fossez}}, \bibinfo {author} {\bibfnamefont {W.}~\bibnamefont {Nazarewicz}},
  \bibinfo {author} {\bibfnamefont {Y.}~\bibnamefont {Jaganathen}}, \bibinfo
  {author} {\bibfnamefont {N.}~\bibnamefont {Michel}}, \ and\ \bibinfo {author}
  {\bibfnamefont {M.}~\bibnamefont {P{\l}oszajczak}},\ }\href
  {http://dx.doi.org/10.1103/PhysRevC.93.011305} {\bibfield  {journal}
  {\bibinfo  {journal} {Phys. Rev. C}\ }\textbf {\bibinfo {volume} {93}},\
  \bibinfo {pages} {011305(R)} (\bibinfo {year}
  {2016}{\natexlab{b}})}\BibitemShut {NoStop}%
\bibitem [{\citenamefont {Berggren}(1968)}]{berggren68_32}%
  \BibitemOpen
  \bibfield  {author} {\bibinfo {author} {\bibfnamefont {T.}~\bibnamefont
  {Berggren}},\ }\href {https://dx.doi.org/10.1016/0375-9474(68)90593-9}
  {\bibfield  {journal} {\bibinfo  {journal} {Nucl. Phys. A}\ }\textbf
  {\bibinfo {volume} {109}},\ \bibinfo {pages} {265} (\bibinfo {year}
  {1968})}\BibitemShut {NoStop}%
\bibitem [{\citenamefont {Berggren}\ and\ \citenamefont
  {Lind}(1993)}]{berggren93_481}%
  \BibitemOpen
  \bibfield  {author} {\bibinfo {author} {\bibfnamefont {T.}~\bibnamefont
  {Berggren}}\ and\ \bibinfo {author} {\bibfnamefont {P.}~\bibnamefont
  {Lind}},\ }\href {https://dx.doi.org/10.1103/PhysRevC.47.768} {\bibfield
  {journal} {\bibinfo  {journal} {Phys. Rev. C}\ }\textbf {\bibinfo {volume}
  {47}},\ \bibinfo {pages} {768} (\bibinfo {year} {1993})}\BibitemShut
  {NoStop}%
\bibitem [{\citenamefont {Kok}(1980)}]{Kok1980}%
  \BibitemOpen
  \bibfield  {author} {\bibinfo {author} {\bibfnamefont {L.~P.}\ \bibnamefont
  {Kok}},\ }\href {\doibase 10.1103/PhysRevLett.45.427} {\bibfield  {journal}
  {\bibinfo  {journal} {Phys. Rev. Lett.}\ }\textbf {\bibinfo {volume} {45}},\
  \bibinfo {pages} {427} (\bibinfo {year} {1980})}\BibitemShut {NoStop}%
\bibitem [{\citenamefont {Mukhamedzhanov}\ \emph {et~al.}(2010)\citenamefont
  {Mukhamedzhanov}, \citenamefont {Irgaziev}, \citenamefont {Goldberg},
  \citenamefont {Orlov},\ and\ \citenamefont {Qazi}}]{mukhamedzhanov10_210}%
  \BibitemOpen
  \bibfield  {author} {\bibinfo {author} {\bibfnamefont {A.~M.}\ \bibnamefont
  {Mukhamedzhanov}}, \bibinfo {author} {\bibfnamefont {B.~F.}\ \bibnamefont
  {Irgaziev}}, \bibinfo {author} {\bibfnamefont {V.~Z.}\ \bibnamefont
  {Goldberg}}, \bibinfo {author} {\bibfnamefont {Y.~V.}\ \bibnamefont {Orlov}},
  \ and\ \bibinfo {author} {\bibfnamefont {I.}~\bibnamefont {Qazi}},\ }\href
  {https://dx.doi.org/10.1103/PhysRevC.81.054314} {\bibfield  {journal}
  {\bibinfo  {journal} {Phys. Rev. C}\ }\textbf {\bibinfo {volume} {81}},\
  \bibinfo {pages} {054314} (\bibinfo {year} {2010})}\BibitemShut {NoStop}%
\bibitem [{\citenamefont {Mukhamedzhanov}\ \emph {et~al.}(2017)\citenamefont
  {Mukhamedzhanov}, \citenamefont {Shubhchintak},\ and\ \citenamefont
  {Bertulani}}]{Mukhamedzhanov2017}%
  \BibitemOpen
  \bibfield  {author} {\bibinfo {author} {\bibfnamefont {A.~M.}\ \bibnamefont
  {Mukhamedzhanov}}, \bibinfo {author} {\bibnamefont {Shubhchintak}}, \ and\
  \bibinfo {author} {\bibfnamefont {C.~A.}\ \bibnamefont {Bertulani}},\ }\href
  {\doibase 10.1103/PhysRevC.96.024623} {\bibfield  {journal} {\bibinfo
  {journal} {Phys. Rev. C}\ }\textbf {\bibinfo {volume} {96}},\ \bibinfo
  {pages} {024623} (\bibinfo {year} {2017})}\BibitemShut {NoStop}%
\bibitem [{\citenamefont {Sofianos}\ \emph {et~al.}(1997)\citenamefont
  {Sofianos}, \citenamefont {Rakityansky},\ and\ \citenamefont
  {Vermaak}}]{Sofianos1997}%
  \BibitemOpen
  \bibfield  {author} {\bibinfo {author} {\bibfnamefont {S.~A.}\ \bibnamefont
  {Sofianos}}, \bibinfo {author} {\bibfnamefont {S.~A.}\ \bibnamefont
  {Rakityansky}}, \ and\ \bibinfo {author} {\bibfnamefont {G.~P.}\ \bibnamefont
  {Vermaak}},\ }\href {http://stacks.iop.org/0954-3899/23/i=11/a=010}
  {\bibfield  {journal} {\bibinfo  {journal} {J. Phys. G}\ }\textbf {\bibinfo
  {volume} {23}},\ \bibinfo {pages} {1619} (\bibinfo {year}
  {1997})}\BibitemShut {NoStop}%
\bibitem [{\citenamefont {Gyarmati}\ and\ \citenamefont
  {Vertse}(1971)}]{gyarmati71_38}%
  \BibitemOpen
  \bibfield  {author} {\bibinfo {author} {\bibfnamefont {B.}~\bibnamefont
  {Gyarmati}}\ and\ \bibinfo {author} {\bibfnamefont {T.}~\bibnamefont
  {Vertse}},\ }\href {https://dx.doi.org/10.1016/0375-9474(71)90095-9}
  {\bibfield  {journal} {\bibinfo  {journal} {Nucl. Phys. A}\ }\textbf
  {\bibinfo {volume} {160}},\ \bibinfo {pages} {523} (\bibinfo {year}
  {1971})}\BibitemShut {NoStop}%
\bibitem [{\citenamefont {Simon}(1979)}]{simon79_436}%
  \BibitemOpen
  \bibfield  {author} {\bibinfo {author} {\bibfnamefont {B.}~\bibnamefont
  {Simon}},\ }\href {https://dx.doi.org/10.1016/0375-9601(79)90165-8}
  {\bibfield  {journal} {\bibinfo  {journal} {Phys. Lett. A}\ }\textbf
  {\bibinfo {volume} {71}},\ \bibinfo {pages} {211} (\bibinfo {year}
  {1979})}\BibitemShut {NoStop}%
\bibitem [{\citenamefont {Neirotti}\ \emph {et~al.}(1997)\citenamefont
  {Neirotti}, \citenamefont {Serra},\ and\ \citenamefont
  {Kais}}]{neirotti97_1466}%
  \BibitemOpen
  \bibfield  {author} {\bibinfo {author} {\bibfnamefont {J.~P.}\ \bibnamefont
  {Neirotti}}, \bibinfo {author} {\bibfnamefont {P.}~\bibnamefont {Serra}}, \
  and\ \bibinfo {author} {\bibfnamefont {S.}~\bibnamefont {Kais}},\ }\href
  {http://dx.doi.org/10.1103/PhysRevLett.79.3142} {\bibfield  {journal}
  {\bibinfo  {journal} {Phys. Rev. Lett.}\ }\textbf {\bibinfo {volume} {79}},\
  \bibinfo {pages} {3142} (\bibinfo {year} {1997})}\BibitemShut {NoStop}%
\bibitem [{\citenamefont {Ferron}\ \emph {et~al.}(2004)\citenamefont {Ferron},
  \citenamefont {Serra},\ and\ \citenamefont {Kais}}]{ferron04_130}%
  \BibitemOpen
  \bibfield  {author} {\bibinfo {author} {\bibfnamefont {A.}~\bibnamefont
  {Ferron}}, \bibinfo {author} {\bibfnamefont {P.}~\bibnamefont {Serra}}, \
  and\ \bibinfo {author} {\bibfnamefont {S.}~\bibnamefont {Kais}},\ }\href
  {https://dx.doi.org/10.1063/1.1695552} {\bibfield  {journal} {\bibinfo
  {journal} {J. Chem. Phys.}\ }\textbf {\bibinfo {volume} {120}},\ \bibinfo
  {pages} {8412} (\bibinfo {year} {2004})}\BibitemShut {NoStop}%
\bibitem [{\citenamefont {Pupyshev}\ and\ \citenamefont
  {Ermilov}(2004)}]{pupyshev04_161}%
  \BibitemOpen
  \bibfield  {author} {\bibinfo {author} {\bibfnamefont {V.~I.}\ \bibnamefont
  {Pupyshev}}\ and\ \bibinfo {author} {\bibfnamefont {A.~Y.}\ \bibnamefont
  {Ermilov}},\ }\href {https://dx.doi.org/10.1002/qua.10570} {\bibfield
  {journal} {\bibinfo  {journal} {Int. J. Quant. Chem.}\ }\textbf {\bibinfo
  {volume} {96}},\ \bibinfo {pages} {185} (\bibinfo {year} {2004})}\BibitemShut
  {NoStop}%
\bibitem [{\citenamefont {Misu}\ \emph {et~al.}(1997)\citenamefont {Misu},
  \citenamefont {Nazarewicz},\ and\ \citenamefont {{\AA}berg}}]{misu97_1181}%
  \BibitemOpen
  \bibfield  {author} {\bibinfo {author} {\bibfnamefont {T.}~\bibnamefont
  {Misu}}, \bibinfo {author} {\bibfnamefont {W.}~\bibnamefont {Nazarewicz}}, \
  and\ \bibinfo {author} {\bibfnamefont {S.}~\bibnamefont {{\AA}berg}},\ }\href
  {https://dx.doi.org/10.1016/S0375-9474(96)00458-7} {\bibfield  {journal}
  {\bibinfo  {journal} {Nucl. Phys. A}\ }\textbf {\bibinfo {volume} {614}},\
  \bibinfo {pages} {44} (\bibinfo {year} {1997})}\BibitemShut {NoStop}%
\bibitem [{\citenamefont {Riisager}\ \emph {et~al.}(1992)\citenamefont
  {Riisager}, \citenamefont {Jensen},\ and\ \citenamefont
  {M{\o}ller}}]{riisager92_615}%
  \BibitemOpen
  \bibfield  {author} {\bibinfo {author} {\bibfnamefont {K.}~\bibnamefont
  {Riisager}}, \bibinfo {author} {\bibfnamefont {A.~S.}\ \bibnamefont
  {Jensen}}, \ and\ \bibinfo {author} {\bibfnamefont {P.}~\bibnamefont
  {M{\o}ller}},\ }\href {https://dx.doi.org/10.1016/0375-9474(92)90691-C}
  {\bibfield  {journal} {\bibinfo  {journal} {Nucl. Phys. A}\ }\textbf
  {\bibinfo {volume} {548}},\ \bibinfo {pages} {393} (\bibinfo {year}
  {1992})}\BibitemShut {NoStop}%
\bibitem [{\citenamefont {Yoshida}\ and\ \citenamefont
  {Hagino}(2005)}]{yoshida05_1895}%
  \BibitemOpen
  \bibfield  {author} {\bibinfo {author} {\bibfnamefont {K.}~\bibnamefont
  {Yoshida}}\ and\ \bibinfo {author} {\bibfnamefont {K.}~\bibnamefont
  {Hagino}},\ }\href {http://dx.doi.org/10.1103/PhysRevC.72.064311} {\bibfield
  {journal} {\bibinfo  {journal} {Phys. Rev. C}\ }\textbf {\bibinfo {volume}
  {72}},\ \bibinfo {pages} {064311} (\bibinfo {year} {2005})}\BibitemShut
  {NoStop}%
\bibitem [{\citenamefont {{Id Betan}}\ \emph {et~al.}(2004)\citenamefont {{Id
  Betan}}, \citenamefont {Liotta}, \citenamefont {Sandulescu},\ and\
  \citenamefont {Vertse}}]{betan04_37}%
  \BibitemOpen
  \bibfield  {author} {\bibinfo {author} {\bibfnamefont {R.~M.}\ \bibnamefont
  {{Id Betan}}}, \bibinfo {author} {\bibfnamefont {R.~J.}\ \bibnamefont
  {Liotta}}, \bibinfo {author} {\bibfnamefont {N.}~\bibnamefont {Sandulescu}},
  \ and\ \bibinfo {author} {\bibfnamefont {T.}~\bibnamefont {Vertse}},\ }\href
  {https://dx.doi.org/10.1016/j.physletb.2004.01.042} {\bibfield  {journal}
  {\bibinfo  {journal} {Phys. Lett. B}\ }\textbf {\bibinfo {volume} {584}},\
  \bibinfo {pages} {48} (\bibinfo {year} {2004})}\BibitemShut {NoStop}%
\bibitem [{\citenamefont {Michel}\ \emph {et~al.}(2006)\citenamefont {Michel},
  \citenamefont {Nazarewicz}, \citenamefont {P{\l}oszajczak},\ and\
  \citenamefont {Rotureau}}]{michel06_16}%
  \BibitemOpen
  \bibfield  {author} {\bibinfo {author} {\bibfnamefont {N.}~\bibnamefont
  {Michel}}, \bibinfo {author} {\bibfnamefont {W.}~\bibnamefont {Nazarewicz}},
  \bibinfo {author} {\bibfnamefont {M.}~\bibnamefont {P{\l}oszajczak}}, \ and\
  \bibinfo {author} {\bibfnamefont {J.}~\bibnamefont {Rotureau}},\ }\href
  {https://dx.doi.org/10.1103/PhysRevC.74.054305} {\bibfield  {journal}
  {\bibinfo  {journal} {Phys. Rev. C}\ }\textbf {\bibinfo {volume} {74}},\
  \bibinfo {pages} {054305} (\bibinfo {year} {2006})}\BibitemShut {NoStop}%
\bibitem [{\citenamefont {Michel}\ \emph {et~al.}(2009)\citenamefont {Michel},
  \citenamefont {Nazarewicz}, \citenamefont {P{\l}oszajczak},\ and\
  \citenamefont {Vertse}}]{michel09_2}%
  \BibitemOpen
  \bibfield  {author} {\bibinfo {author} {\bibfnamefont {N.}~\bibnamefont
  {Michel}}, \bibinfo {author} {\bibfnamefont {W.}~\bibnamefont {Nazarewicz}},
  \bibinfo {author} {\bibfnamefont {M.}~\bibnamefont {P{\l}oszajczak}}, \ and\
  \bibinfo {author} {\bibfnamefont {T.}~\bibnamefont {Vertse}},\ }\href
  {https://dx.doi.org/10.1088/0954-3899/36/1/013101} {\bibfield  {journal}
  {\bibinfo  {journal} {J. Phys. G}\ }\textbf {\bibinfo {volume} {36}},\
  \bibinfo {pages} {013101} (\bibinfo {year} {2009})}\BibitemShut {NoStop}%
\bibitem [{\citenamefont {Rohr}\ and\ \citenamefont
  {Linder}(1975)}]{rohr75_283}%
  \BibitemOpen
  \bibfield  {author} {\bibinfo {author} {\bibfnamefont {K.}~\bibnamefont
  {Rohr}}\ and\ \bibinfo {author} {\bibfnamefont {F.}~\bibnamefont {Linder}},\
  }\href {https://dx.doi.org/10.1088/0022-3700/8/10/009} {\bibfield  {journal}
  {\bibinfo  {journal} {J. Phys. B}\ }\textbf {\bibinfo {volume} {8}},\
  \bibinfo {pages} {L200} (\bibinfo {year} {1975})}\BibitemShut {NoStop}%
\bibitem [{\citenamefont {Rohr}\ and\ \citenamefont
  {Linder}(1976)}]{rohr76_284}%
  \BibitemOpen
  \bibfield  {author} {\bibinfo {author} {\bibfnamefont {K.}~\bibnamefont
  {Rohr}}\ and\ \bibinfo {author} {\bibfnamefont {F.}~\bibnamefont {Linder}},\
  }\href {https://dx.doi.org/10.1088/0022-3700/9/14/020} {\bibfield  {journal}
  {\bibinfo  {journal} {J. Phys. B}\ }\textbf {\bibinfo {volume} {9}},\
  \bibinfo {pages} {2521} (\bibinfo {year} {1976})}\BibitemShut {NoStop}%
\bibitem [{\citenamefont {Rohr}(1978)}]{rohr78_1426}%
  \BibitemOpen
  \bibfield  {author} {\bibinfo {author} {\bibfnamefont {K.}~\bibnamefont
  {Rohr}},\ }\href {http://dx.doi.org/10.1088/0022-3700/11/10/019} {\bibfield
  {journal} {\bibinfo  {journal} {J. Phys. B}\ }\textbf {\bibinfo {volume}
  {11}},\ \bibinfo {pages} {1849} (\bibinfo {year} {1978})}\BibitemShut
  {NoStop}%
\bibitem [{\citenamefont {Heiss}\ and\ \citenamefont
  {Nazmitdinov}(2011)}]{heiss11_776}%
  \BibitemOpen
  \bibfield  {author} {\bibinfo {author} {\bibfnamefont {W.~D.}\ \bibnamefont
  {Heiss}}\ and\ \bibinfo {author} {\bibfnamefont {R.~G.}\ \bibnamefont
  {Nazmitdinov}},\ }\href {https://dx.doi.org/10.1140/epjd/e2011-20174-4}
  {\bibfield  {journal} {\bibinfo  {journal} {Eur. Phys. J. D.}\ }\textbf
  {\bibinfo {volume} {63}},\ \bibinfo {pages} {369} (\bibinfo {year}
  {2011})}\BibitemShut {NoStop}%
\bibitem [{\citenamefont {Heiss}(2012)}]{heiss12_1392}%
  \BibitemOpen
  \bibfield  {author} {\bibinfo {author} {\bibfnamefont {W.~D.}\ \bibnamefont
  {Heiss}},\ }\href {http://dx.doi.org/10.1088/1751-8113/45/44/444016}
  {\bibfield  {journal} {\bibinfo  {journal} {J. Phys. A}\ }\textbf {\bibinfo
  {volume} {45}},\ \bibinfo {pages} {444016} (\bibinfo {year}
  {2012})}\BibitemShut {NoStop}%
\bibitem [{\citenamefont {M\"uller}\ and\ \citenamefont
  {Rotter}(2008)}]{Muller2008_1973}%
  \BibitemOpen
  \bibfield  {author} {\bibinfo {author} {\bibfnamefont {M.}~\bibnamefont
  {M\"uller}}\ and\ \bibinfo {author} {\bibfnamefont {I.}~\bibnamefont
  {Rotter}},\ }\href {http://stacks.iop.org/1751-8121/41/i=24/a=244018}
  {\bibfield  {journal} {\bibinfo  {journal} {J. Phys. A}\ }\textbf {\bibinfo
  {volume} {41}},\ \bibinfo {pages} {244018} (\bibinfo {year}
  {2008})}\BibitemShut {NoStop}%
\bibitem [{\citenamefont {Oko\l{}owicz}\ and\ \citenamefont
  {P\l{}oszajczak}(2009)}]{Okolowicz2009_1974}%
  \BibitemOpen
  \bibfield  {author} {\bibinfo {author} {\bibfnamefont {J.}~\bibnamefont
  {Oko\l{}owicz}}\ and\ \bibinfo {author} {\bibfnamefont {M.}~\bibnamefont
  {P\l{}oszajczak}},\ }\href {\doibase 10.1103/PhysRevC.80.034619} {\bibfield
  {journal} {\bibinfo  {journal} {Phys. Rev. C}\ }\textbf {\bibinfo {volume}
  {80}},\ \bibinfo {pages} {034619} (\bibinfo {year} {2009})}\BibitemShut
  {NoStop}%
\bibitem [{\citenamefont {Domcke}(1981)}]{domcke81_289}%
  \BibitemOpen
  \bibfield  {author} {\bibinfo {author} {\bibfnamefont {W.}~\bibnamefont
  {Domcke}},\ }\href {https://dx.doi.org/10.1088/0022-3700/14/24/022}
  {\bibfield  {journal} {\bibinfo  {journal} {J. Phys. B}\ }\textbf {\bibinfo
  {volume} {14}},\ \bibinfo {pages} {4889} (\bibinfo {year}
  {1981})}\BibitemShut {NoStop}%
\bibitem [{\citenamefont {Garmon}\ \emph {et~al.}(2015)\citenamefont {Garmon},
  \citenamefont {Gianfreda},\ and\ \citenamefont {Hatano}}]{Garmon2015_1975}%
  \BibitemOpen
  \bibfield  {author} {\bibinfo {author} {\bibfnamefont {S.}~\bibnamefont
  {Garmon}}, \bibinfo {author} {\bibfnamefont {M.}~\bibnamefont {Gianfreda}}, \
  and\ \bibinfo {author} {\bibfnamefont {N.}~\bibnamefont {Hatano}},\ }\href
  {\doibase 10.1103/PhysRevA.92.022125} {\bibfield  {journal} {\bibinfo
  {journal} {Phys. Rev. A}\ }\textbf {\bibinfo {volume} {92}},\ \bibinfo
  {pages} {022125} (\bibinfo {year} {2015})}\BibitemShut {NoStop}%
\end{thebibliography}

%

\end{document}